\documentclass[twocolumn]{aastex6}
\usepackage{natbib}
\usepackage{amsmath}
\usepackage{graphicx}
\renewcommand{\phi}{\varphi}
\newcommand{\rhessi}{\emph{RHESSI}\ }
\newcommand{\goes}{\emph{GOES}\ }
\usepackage{todonotes}

\shorttitle{Hard X-Ray Emission from Partially Occulted Solar Flares}
\shortauthors{Effenberger et al.}

\begin{document}

\received{?}
\accepted{?}

\title{Hard X-Ray Emission from Partially Occulted Solar Flares:
  RHESSI Observations in Two Solar Cycles}

\author{Frederic~Effenberger} \and
\author{Fatima~Rubio~da~Costa} \affil{Department of Physics and KIPAC,
  Stanford University, Stanford, CA 94305, USA}
\email{feffen@stanford.edu, frubio@stanford.edu} \and
\author{Mitsuo~Oka} \and
\author{Pascal~Saint-Hilaire} \affil{Space Sciences Laboratory,
  University of California, Berkeley, CA 94720-7450, USA}\and
\author{Wei~Liu} \affil{Bay Area Environmental Research Institute, 625
  2nd Street, Suite 209, Petaluma, CA 94952, USA \\Lockheed Martin
  Solar and Astrophysics Laboratory, 3251 Hanover Street, Bldg.~252,
  Palo Alto, CA 94304, USA\\W.~W.~Hansen Experimental Physics
  Laboratory, Stanford University, Stanford, CA 94305, USA}\and
\author{Vah\'e~Petrosian} \affil{Department of Physics and KIPAC,
  Stanford University, Stanford, CA 94305, USA} \and
\author{Lindsay~Glesener} \affil{School of Physics and Astronomy, University of Minnesota, Minneapolis, MN 55455, USA} \and
\author{S\"am~Krucker} \affil{Space Sciences Laboratory, University of California,
Berkeley, CA 94720-7450, USA\\ University of Applied Sciences and Arts Northwestern Switzerland, Bahnhofstrasse 6, 5210 Windisch, Switzerland}

\begin{abstract}
  Flares close to the solar limb, where the footpoints are occulted,
  can reveal the spectrum and structure of the coronal loop-top source
  in X-rays. We aim at studying the properties of the corresponding
  energetic electrons near their acceleration site, without footpoint
  contamination. To this end, a statistical study of partially
  occulted flares observed with \emph{RHESSI} is presented here,
  covering a large part of solar cycles 23 and 24. We perform a
  detailed spectra, imaging and light curve analysis for 116 flares
  and include contextual observations from \emph{SDO} and
  \emph{STEREO} when available, providing further insights into flare
  emission that was previously not accessible. We find that most
  spectra are fitted well with a thermal component plus a broken
  power-law, non-thermal component. A thin-target kappa distribution
  model gives satisfactory fits after the addition of a thermal
  component. X-rays imaging reveals small spatial separation between
  the thermal and non-thermal components, except for a few flares with
  a richer coronal source structure. A comprehensive light curve
  analysis shows a very good correlation between the derivative of the
  soft X-ray flux (from \emph{GOES}) and the hard X-rays for a
  substantial number of flares, indicative of the Neupert effect. The
  results confirm that non-thermal particles are accelerated in the
  corona and estimated timescales support the validity of a
  thin-target scenario with similar magnitudes of thermal and
  non-thermal energy fluxes.
\end{abstract}


\section{Introduction}
Electron transport and acceleration in solar flares are a major topic
in contemporary high-energy solar flare research. The main
observational tool in these investigations are hard X-ray emissions
(mainly non-thermal bremsstrahlung) emitted by the energetic electron
distribution. The {\it Reuven Ramaty High-Energy Solar Spectroscopic
  Imager} \citep[\emph{RHESSI},][]{Lin-etal-2002} is a unique
instrument to reveal the spectral and spatial properties of these
emissions.

Studies in the past have shown that in many flares at least two
distinct types of sources can be distinguished, namely from the
coronal solar flare loop-top and from chromospheric footpoints
\citep{Masuda-etal-1994, Petrosian-etal-2002, Krucker-etal-2007,
  Simoes-Kontar-2013}. Theories suggest \citep[see, e.g., the review
by][]{Petrosian-2012} that the coronal region at the loop-top is the
main acceleration site for electrons, however, due to the limited
dynamical range of \emph{RHESSI}, it is often hard to clearly
observe coronal sources, when strong footpoint emission is
present. Partially occulted flares, in which the footpoints are behind
the solar limb, offer the opportunity to observe the coronal sources
in isolation. \citet{Krucker-Lin-2008} (hereafter KL2008) studied a
selection of 55 partially occulted flares from March 2002 to August
2004 covering the maximum of solar cycle 23 with \emph{RHESSI}. They
found that the photon spectra at high-energies show a steep (soft)
spectral index (mostly between 4 and 6) and concluded that thin-target
emission in the corona from flare-accelerated electrons is consistent
with the observations.

Previous studies of partially occulted flares involved also data from
the \emph{Yohkoh} mission \citep{Tomczak-2009}.
\citet{Bai-etal-2012} investigated an extended list including \emph{RHESSI}
flares until the end of 2010, however, the deep solar minimum prevented
a substantial extension of the KL2008 selection. Recently,
observations with the \emph{Fermi} Large Area Telescope (LAT) of
behind-the-limb flares in gamma rays \citep{Pesce-Rollins-etal-2015}
sparked additional interest in occulted flares and coronal sources
\citep[see also][for an earlier event study]{Vilmer-etal-1999}. In
particular the question of confinement of the energetic particle
population near the acceleration region in the corona is a central
issue; see for example the modeling studies of
\citet{Kontar-etal-2014} and the observations discussed in
\citet{Simoes-Kontar-2013} and \citet{Chen-Petrosian-2013}.

The coronal sources sometimes show a rich morphology, with emission above
and below the presumable reconnection region
\citep[e.g.][]{Liu-etal-2013}. Separated sources have for example been
analyzed by \citet{Battaglia-Benz-2006} and are also of interest in the
context of novel modeling approaches with kappa functions
\citep{Bian-etal-2014, Oka-etal-2013}. We thus systematically include a
thermal plus thin-target kappa function fit in our analysis, as first
introduced by \citet[][]{Kasparova-Karlicky-2009}.  For further
details on observational and modeling aspects of coronal sources we
point to the review by \citet{Krucker-etal-2008}.

To improve on our knowledge of coronal source properties and the
associated non-thermal electrons, a detailed spectra, imaging and
light curve analysis for 116 partially occulted flares is performed in
this study, covering large parts of solar cycles 23 and 24. For the
first time, we systematically include contextual observations from
\emph{SDO} and \emph{STEREO} when available, to provide further
insights into flare emission that was previously not
accessible. Additionally, we present a comprehensive light curve
analysis between the derivative of the soft X-ray flux (from
\emph{GOES}) and the hard X-rays for a substantial number of flares,
indicative of the so-called Neupert effect \citep[]{Neupert-1968}.

We introduce the data analysis methods and partially occulted flare
sample studied here in Section~\ref{Sect:analysis}, together with an
overview of the results and their statistics. Further analysis and
discussion of the results is presented in
Section~\ref{Sect:discussion} followed by a summary. The Appendix
contains the results obtained for the 55 KL2008 flares, applying our
methodology.

%%%%%%%%%%%%%%%%%%%%%%%%%%%%%%%%%%%%%%%%%%%%%%%%%%%%%%%%

\section{Data analysis and statistics of partially occulted flares}
\begin{figure}[ht]
\centering
\noindent\includegraphics[width=0.49\textwidth]{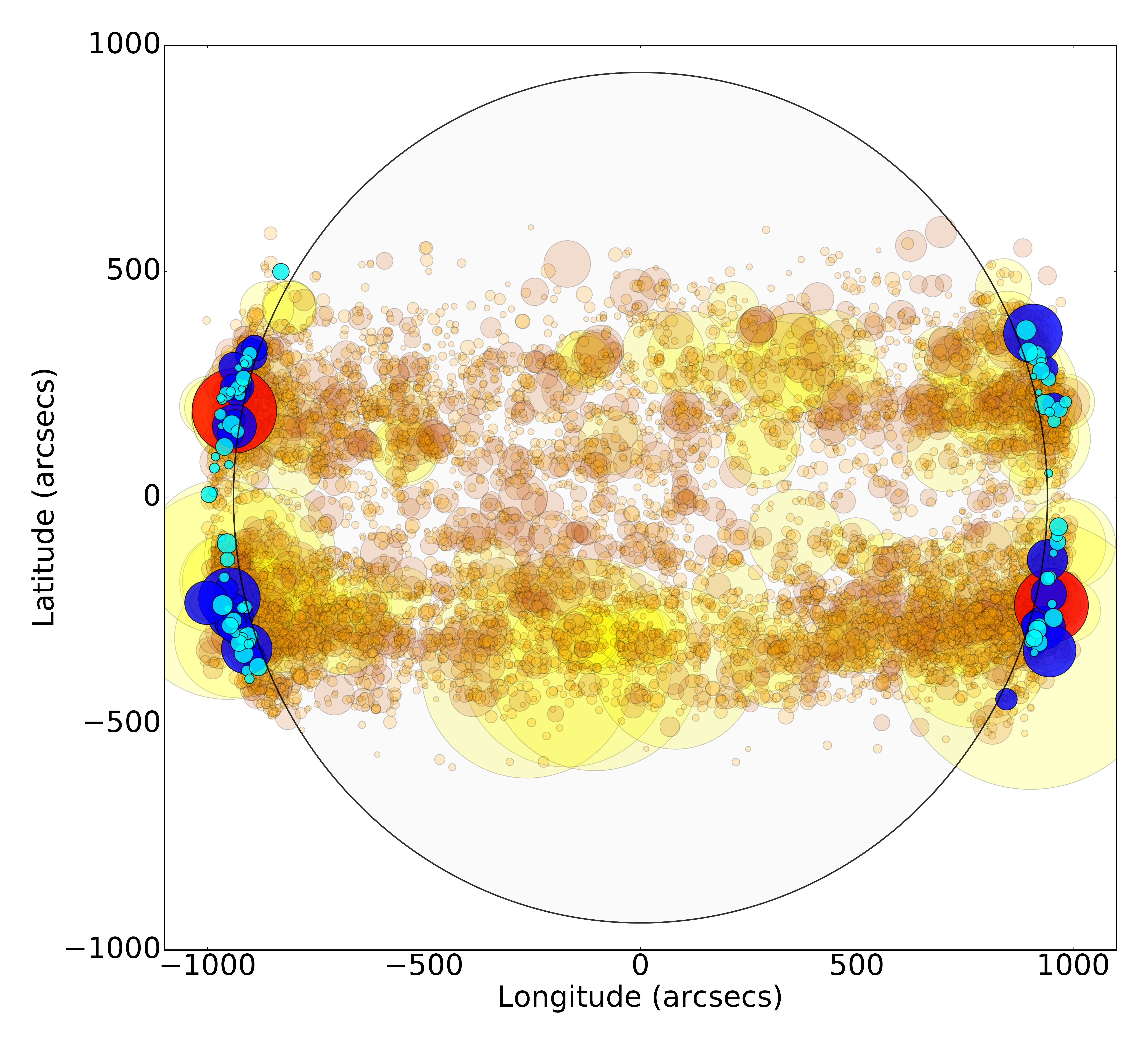}
\caption{Positions of the partially occulted flares
  selected in this study (C-Class: light green; M-Class: dark blue;
  X-Class: red) in context of all C-Class and above flares
  from the \rhessi flare catalog until the end of 2015 (yellow and
  orange). The size of the circles is scaled with the observed \goes
  level and the solar limb is drawn at 940~arcsecs to guide the eye. We
  omitted outliers with Solar-Y greater or smaller than $\pm 600$
  arcsecs, as being unphysical.}
\label{fig:disk}
\end{figure}
For the purpose of this study we extended the list of partially
occulted flare candidates to cover solar cycles 23 and 24. Our study
is based on two joined data sets. For the time interval from March
2002 until August 2004 we used the same selection of flares as
discussed in KL2008 and included them in our analysis. The
second data set is based on partially occulted flare candidates in the
\emph{RHESSI} flare catalog, covering flares simultaneously observed
by \emph{SDO}, from January 2011 until December 2015.\footnote{The list
  of candidate flares with additional information can be accessed on
  the web at:
  \url{http://www.ssl.berkeley.edu/~moka/rhessi/flares_occulted.html}}

\subsection{Flare selection}
The candidate flares with occulted footpoints were selected from the
\emph{RHESSI} flare list as those flares with significant counts at
energies of 25~keV and higher and being close to the solar limb
(centroid position $\sim930-1050$~arcsec with respect to the solar
center).

Using {\it SDO}/AIA, \emph{STEREO} and \emph{RHESSI}, we inspected the
emission in the hot corona and X-rays of approximately 400 candidates
to visually determine which ones are actually occulted in their
footpoints. We aimed to avoid false-positives, i.e. flares which are
not truly occulted, as much as possible in our selection, to prevent
contamination of footpoint emission in our spectral
analysis. As such, this selection can be regarded as a conservative
lower limit approximation to all the actually partially occulted
flares observed by {\it RHESSI}. We nevertheless expect no significant
biases in our selection, but the overall sample size has to be
kept in mind.

Table~\ref{occulted_list} gives a list of 61 flares from solar cycle
24 satisfying our selection criteria with date, time, {\it GOES}
classification (directly measured and estimated with \emph{STEREO}),
and longitudinal/latitudinal centroid position in the higher energy
range according to the \emph{RHESSI} flare list. We also list the main
parameters resulting from our spectral, light-curve and imaging
analysis, as described in the following sections (the results for the
55 KL2008 flares are listed in Table~\ref{KL_list}).\footnote{The
  results are available as csv files together with a python notebook
  containing the analysis and figures; at GitHub:
  \url{https://github.com/feffenberger/occulted-flares} and in the Stanford
  Digital Repository: \url{https://purl.stanford.edu/fp125hq3736}}

Figure~\ref{fig:disk} gives an overview of the positions of our
selection of flares, in the context of all flares observed by
\emph{RHESSI} (C-class and above). The circle size is scaled
proportionally to the observed {\it GOES} class.

\label{Sect:analysis}
\begin{deluxetable*}{rllccrrc|ccc|ccc|c|cr}
\centering
\tabletypesize{\scriptsize}
\tablecolumns{17}
\tablewidth{0pt}  
\tablecaption{\label{occulted_list} Partially occulted flares from solar cycle
  24. See Table~\ref{KL_list} in the Appendix for the flare list from
  the previous cycle.}
\tablehead{
  \colhead{\#} & 
  \colhead{Date} & 
  \colhead{Time} &
  \multicolumn{2}{c}{{\it GOES} Class} & 
  \colhead{Sol-X} &
  \colhead{Sol-Y} &
  \colhead{$H$} &
  \colhead{$T_{th}$} &
  \colhead{$E_{break}$} &
  \colhead{$\gamma$} &
  \colhead{$T_{th}^{\kappa}$} &
  \colhead{$T^{\kappa}$} &
  \colhead{$\kappa$} &
  \colhead{$d_{max}$\tablenotemark{(a)}} &
  \colhead{Lin.}  &
  \colhead{Lag \tablenotemark{(b)}}\vspace{0.1cm}\\
  \colhead{} &
  \colhead{} &
  \colhead{(UT)} &
  \colhead{Orig.} &
  \colhead{{\it STER.}} &
  \colhead{(arcs.)} &
  \colhead{(arcs.)} &
  \colhead{(Mm)} &
  \colhead{(MK)} &
  \colhead{(keV)} &      
  \colhead{} &
  \colhead{(MK)} &
  \colhead{(MK)} &
  \colhead{} &
  \colhead{(Mm)} &
  \colhead{Corr.} &
  \colhead{(s)}\vspace{0.1cm}\\
  \colhead{1} &
  \colhead{2} &
  \colhead{3} &
  \colhead{4} &
  \colhead{5} &
  \colhead{6} &
  \colhead{7} &
  \colhead{8} &
  \colhead{9} &
  \colhead{10} &      
  \colhead{11} &
  \colhead{12} &
  \colhead{13} &
  \colhead{14} &
  \colhead{15} &
  \colhead{16} &
  \colhead{17}}
\startdata
\setlength{\tabcolsep}{0.1pt}
1  &  2011 Jan 28 &  00:57:03 &  M1.3 &   M9.3 &     937.7 &    285.6 &         3.7 &      20.1 &        15.6 &     4.73 &               18.2 &         12.9 &     4.33 &           -1.2 &        0.81 &   0 \\
2  &  2011 Mar 08 &  18:12:43 &  M4.4 &   M7.6 &     932.0 &   -287.6 &         6.3 &      25.0 &        18.4 &     4.68 &               18.0 &         12.4 &     4.30 &           -1.6 &        0.93 &   0 \\
3  &  2011 Jun 12 &  17:07:47 &  B8.6 &   C2.4 &    -915.7 &    275.3 &        16.6 &      19.2 &        13.1 &     7.74 &                - &          - &      - &            0.5 &         - & - \\
4  &  2011 Sep 06 &  06:00:25 &  C9.6 &   X1.7 &     890.7 &    370.6 &        24.4 &      19.6 &        16.3 &     7.32 &               12.9 &         14.0 &     7.90 &           -0.8 &         - & - \\
5  &  2012 Aug 17 &  08:33:04 &  C4.7 &   C9.5 &    -902.2 &    318.4 &         3.7 &      22.2 &        18.0 &     3.75 &               18.5 &          8.0 &     2.87 &            1.9 &        0.72 & -16 \\
6  &  2012 Aug 17 &  13:17:50 &  M2.4 &   M3.9 &    -898.8 &    316.9 &         3.7 &      24.4 &        21.4 &     5.03 &               22.0 &          5.6 &     5.42 &            1.1 &        0.85 &   0 \\
7  &  2012 Sep 30 &  23:38:36 &  C9.9 &   M2.8 &     934.4 &    206.0 &         5.7 &      25.2 &         - &      - &                - &          - &      - &            0.8 &         - & - \\
8  &  2012 Oct 07 &  20:24:48 &  C1.2 &   M6.0 &    -928.6 &    287.6 &        18.2 &      34.2 &        12.3 &     4.13 &               21.2 &         26.8 &     5.70 &            0.6 &         - & - \\
9  &  2012 Oct 17 &  07:53:55 &  C7.4 &   M2.1 &    -960.8 &    112.9 &         7.4 &      31.0 &        11.3 &     4.11 &               19.2 &         19.1 &     4.30 &           -0.8 &         - & - \\
10 &  2012 Oct 20 &  18:12:12 &  M9.0 &   M7.1 &    -949.4 &   -222.9 &         7.9 &      24.8 &        18.7 &     7.63 &               28.1 &         15.8 &     7.39 &            0.9 &        0.86 &   8 \\
11 &  2013 Apr 11 &  22:50:23 &  C4.0 &   M2.6 &     955.7 &    169.5 &        15.1 &      25.2 &        17.6 &     6.36 &               22.0 &          9.9 &     5.10 &           -2.1 &         - & - \\
12 &  2013 May 12 &  22:41:06 &  M1.3 &   C5.4 &    -945.4 &    167.3 &        12.6 &      26.6 &        19.5 &     6.84 &               22.7 &         19.9 &     7.54 &            0.1 &         - & - \\
13 &  2013 May 13 &  01:59:15 &  X1.7 &   X1.5 &    -938.0 &    192.2 &        13.3 &      28.6 &        19.9 &     5.79 &               16.1 &         21.6 &     7.31 &            0.9 &        0.70 &   4 \\
14 &  2013 Jul 29 &  23:25:48 &  C6.3 &   C9.3 &     963.5 &    -97.8 &         - &      17.6 &        13.5 &     8.38 &                - &          - &      - &            - &         - & - \\
15 &  2013 Aug 22 &  05:11:32 &  C3.3 &   C7.6 &     962.3 &    -83.8 &         7.5 &      20.2 &         - &      - &                - &          - &      - &            - &         - & - \\
16 &  2013 Oct 14 &  21:46:13 &  C3.0 &   C6.9 &    -970.4 &    184.1 &        20.1 &      22.9 &         - &      - &                - &          - &      - &            - &         - & - \\
17 &  2013 Nov 15 &  10:00:08 &  C1.8 &   M2.2 &    -926.2 &    253.7 &        24.1 &      23.6 &         - &      - &                - &          - &      - &            - &         - & - \\
18 &  2013 Dec 25 &  18:56:52 &  B9.3 &   C1.0 &     965.4 &   -285.4 &         8.8 &      23.7 &         - &      - &                - &          - &      - &            - &         - & - \\
19 &  2013 Dec 31 &  04:45:44 &  C1.4 &   C3.7 &    -967.3 &    -89.2 &        24.0 &      25.9 &         - &      - &                - &          - &      - &            - &         - & - \\
20 &  2014 Jan 16 &  15:18:13 &  C2.8 &    - &    -909.0 &   -382.4 &        26.1 &      18.7 &        12.6 &     6.70 &                - &          - &      - &           -2.1 &         - & - \\
21 &  2014 Jan 17 &  08:28:12 &  C2.2 &    - &    -903.1 &   -400.3 &         9.4 &      19.7 &        12.9 &     6.46 &               21.9 &          6.5 &     4.74 &            0.9 &         - & - \\
22 &  2014 Jan 27 &  22:09:24 &  M4.9 &   M3.1 &    -946.8 &   -260.6 &         1.9 &      14.8 &        14.8 &     6.13 &                7.5 &          8.7 &     6.02 &            1.4 &        0.77 &   0 \\
23 &  2014 Mar 14 &  10:09:44 &  C5.0 &   M1.4 &     942.5 &    262.7 &         9.4 &      27.1 &         - &      - &                - &          - &      - &            - &        0.39 &  24 \\
24 &  2014 Apr 23 &  08:38:08 &  C1.6 &   C6.4 &     943.0 &     54.3 &        14.1 &      33.1 &         - &      - &                - &          - &      - &            - &         - & - \\
25 &  2014 Apr 25 &  00:20:36 &  X1.3 &   X2.6 &     949.3 &   -237.9 &        11.3 &      13.8 &        14.2 &     3.68 &               10.1 &          4.3 &     2.70 &            4.4 &        0.81 & -12 \\
26 &  2014 May 07 &  06:27:26 &  C3.6 &   M2.2 &     946.3 &   -175.6 &        11.0 &      17.2 &        13.2 &     5.92 &               17.1 &          7.4 &     6.03 &           -0.8 &         - & - \\
27 &  2014 Jun 08 &  09:52:08 &  C2.0 &    - &    -920.3 &   -298.3 &        37.7 &      22.2 &         - &      - &                - &          - &      - &            - &         - & - \\
28 &  2014 Sep 03 &  13:35:36 &  M2.5 &   M1.7 &    -940.6 &   -269.9 &        19.1 &      22.0 &        17.1 &     5.88 &               21.2 &         15.8 &     7.06 &            2.8 &         - & - \\
29 &  2014 Sep 11 &  15:23:37 &  M2.1 &   M4.3 &    -926.7 &    245.7 &         4.4 &      26.5 &        19.3 &     3.22 &               26.2 &          6.5 &     2.28 &            0.9 &        0.78 & -12 \\
30 &  2014 Sep 11 &  21:25:02 &  M1.4 &   M1.7 &    -925.9 &    246.1 &         4.0 &      23.0 &        18.1 &     4.67 &               18.2 &         11.2 &     4.19 &           -0.7 &        0.77 &  -4 \\
31 &  2014 Oct 02 &  22:53:43 &  C3.8 &   -  &     921.5 &   -280.3 &         - &      19.4 &        16.6 &     4.59 &               15.1 &          4.2 &     4.14 &           -1.9 &        0.34 &  20 \\
32 &  2014 Oct 22 &  15:52:49 &  M1.4 &    - &    -957.1 &   -203.3 &         - &      20.0 &        17.5 &     4.01 &               19.2 &          6.7 &     3.39 &            4.0 &         - & - \\
33 &  2014 Oct 31 &  00:35:05 &  C8.2 &    - &     954.2 &   -265.8 &         - &      30.1 &        19.6 &     6.59 &               30.7 &         12.6 &     5.81 &           -6.2 &         - & - \\
34 &  2014 Nov 03 &  11:37:00 &  M2.2 &    - &    -938.6 &    288.1 &         - &      19.9 &        16.6 &     5.32 &               17.0 &         10.6 &     5.20 &            2.2 &         - & - \\
35 &  2014 Dec 25 &  08:46:57 &  C1.9 &    - &     950.8 &   -234.7 &         - &      34.5 &        17.2 &     4.50 &               30.1 &          9.5 &     3.47 &           -0.2 &        0.63 &  16 \\
36 &  2015 Mar 03 &  01:32:05 &  M8.2 &    - &     906.5 &    363.0 &         - &      25.6 &        18.8 &     6.76 &               28.9 &         10.4 &     3.76 &           -0.1 &        0.95 &   0 \\
37 &  2015 Mar 21 &  00:14:52 &  C1.4 &    - &     909.5 &   -342.7 &         - &      34.3 &         - &      - &                - &          - &      - &            - &        0.47 &  16 \\
38 &  2015 Mar 30 &  19:29:56 &  C1.0 &    - &     916.4 &    321.6 &         - &      26.2 &         - &      - &                - &          - &      - &            - &         - & - \\
39 &  2015 Apr 13 &  04:07:49 &  C4.3 &    - &    -910.9 &    300.7 &         - &      24.4 &        18.8 &     4.47 &               21.7 &          6.6 &     3.63 &           -2.5 &        0.84 &   4 \\
40 &  2015 Apr 23 &  02:08:29 &  C2.2 &    - &     945.2 &    189.2 &         - &      17.9 &        13.3 &     7.53 &                - &          - &      - &            - &         - & - \\
41 &  2015 May 04 &  02:53:30 &  C3.0 &    - &    -925.4 &    227.1 &         - &      33.2 &        19.1 &     5.78 &               11.0 &         14.7 &     5.29 &            0.4 &        0.91 &   0 \\
42 &  2015 May 04 &  17:01:57 &  C5.1 &    - &    -930.0 &    242.6 &         - &      30.6 &        23.2 &     4.33 &               24.0 &         14.1 &     5.54 &           -3.9 &        0.57 &   0 \\
43 &  2015 Jun 09 &  18:52:09 &  B7.7 &    - &    -896.5 &   -325.3 &         - &      24.3 &         - &      - &                - &          - &      - &            - &         - & - \\
44 &  2015 Jun 10 &  21:25:56 &  C1.5 &    - &    -898.3 &   -320.0 &         - &      25.0 &         - &      - &                - &          - &      - &            - &         - & - \\
45 &  2015 Jun 11 &  18:04:46 &  C1.8 &    - &    -950.7 &     73.2 &         - &      26.9 &        13.9 &     3.43 &               13.9 &          6.2 &     3.76 &            0.6 &         - & - \\
46 &  2015 Jun 15 &  00:46:23 &  C1.0 &    - &     919.8 &    233.0 &         - &      23.7 &         - &      - &                - &          - &      - &            - &         - & - \\
47 &  2015 Jun 28 &  17:12:28 &  C1.9 &    - &    -914.6 &    250.6 &         - &      19.7 &        13.4 &     8.48 &                - &          - &      - &            - &         - & - \\
48 &  2015 Jul 14 &  12:06:25 &  C1.2 &    - &    -920.3 &    241.2 &         - &      26.4 &         - &      - &                - &          - &      - &            - &         - & - \\
49 &  2015 Oct 04 &  02:38:37 &  M1.0 &    - &     914.7 &   -323.9 &         - &      21.3 &        16.4 &     5.50 &               16.2 &         15.0 &     5.86 &           -1.2 &        0.61 &   0 \\
50 &  2015 Oct 16 &  21:56:44 &  C1.1 &    - &    -968.2 &    158.5 &         - &      23.3 &         - &      - &                - &          - &      - &            - &         - & - \\
51 &  2015 Oct 17 &  01:23:24 &  C3.4 &    - &    -920.5 &   -311.3 &         - &      27.1 &         - &      - &                - &          - &      - &            - &         - & - \\
52 &  2015 Oct 17 &  18:35:57 &  C8.6 &    - &    -916.5 &   -344.7 &         - &      25.6 &         - &      - &                - &          - &      - &            - &         - & - \\
53 &  2015 Oct 17 &  23:16:22 &  C6.6 &    - &    -927.4 &   -321.3 &         - &      22.7 &         - &      - &                - &          - &      - &            - &         - & - \\
54 &  2015 Oct 30 &  14:46:09 &  C3.4 &    - &     982.1 &    211.8 &         - &      10.7 &        10.0 &     3.80 &                - &          - &      - &           -3.1 &         - & - \\
55 &  2015 Dec 09 &  10:53:48 &  C1.2 &   C4.4 &    -953.5 &   -245.5 &         - &      24.7 &         - &      - &                - &          - &      - &            - &         - & - \\
56 &  2015 Dec 12 &  05:06:57 &  C4.9 &   M1.1 &    -956.2 &    229.9 &        10.0 &      34.6 &         - &      - &                - &          - &      - &            - &        0.63 &  16 \\
57 &  2015 Dec 12 &  11:42:48 &  C2.2 &   C2.7 &    -966.1 &    216.8 &         9.5 &      32.4 &         - &      - &                - &          - &      - &            - &        0.58 &   8 \\
58 &  2015 Dec 19 &  13:03:35 &  C1.7 &   M8.2 &    -981.7 &     90.6 &        51.7 &      20.6 &        12.4 &     4.35 &                - &          - &      - &            1.3 &         - & - \\
59 &  2015 Dec 20 &  05:01:20 &  C2.4 &   M1.2 &    -983.9 &     65.5 &        17.3 &      18.4 &         - &      - &                - &          - &      - &            - &         - & - \\
60 &  2015 Dec 20 &  12:40:09 &  C2.0 &    - &    -986.4 &     12.4 &        17.1 &      21.9 &         - &      - &                - &          - &      - &            - &         - & - \\
61 &  2015 Dec 20 &  22:37:00 &  C6.1 &   M1.1 &    -996.8 &      6.9 &        15.0 &      22.9 &        12.7 &     7.36 &                - &          - &      - &           -4.9 &        0.48 &   0 \\
\enddata
\tablenotetext{(a)}{\, A positive $d_{max}$ implies a high-energy source at
greater radial distance.}
\tablenotetext{(b)}{\, Positive lags indicate a delay in the \emph{RHESSI}
  light curve with respect to the \emph{GOES} soft X-ray derivative.}
\end{deluxetable*}

\cleardoublepage
\subsection{Time profiles}
We analyzed the time evolution of the hard X-ray flux measured by
\emph{RHESSI} and compared it with the temporal derivative of the soft
X-ray flux measured by {\it GOES} in both the high ($0.5-4$~\AA) and
low ($1-8$~\AA) energy channels for all selected events. Focusing on
higher energy {\it RHESSI} emission, we calculated the linear
correlation between soft and hard X-rays \citep[the so-called
Neupert effect,][]{Neupert-1968}.

\begin{figure}[ht]
\centering
\includegraphics[width=0.48\textwidth]{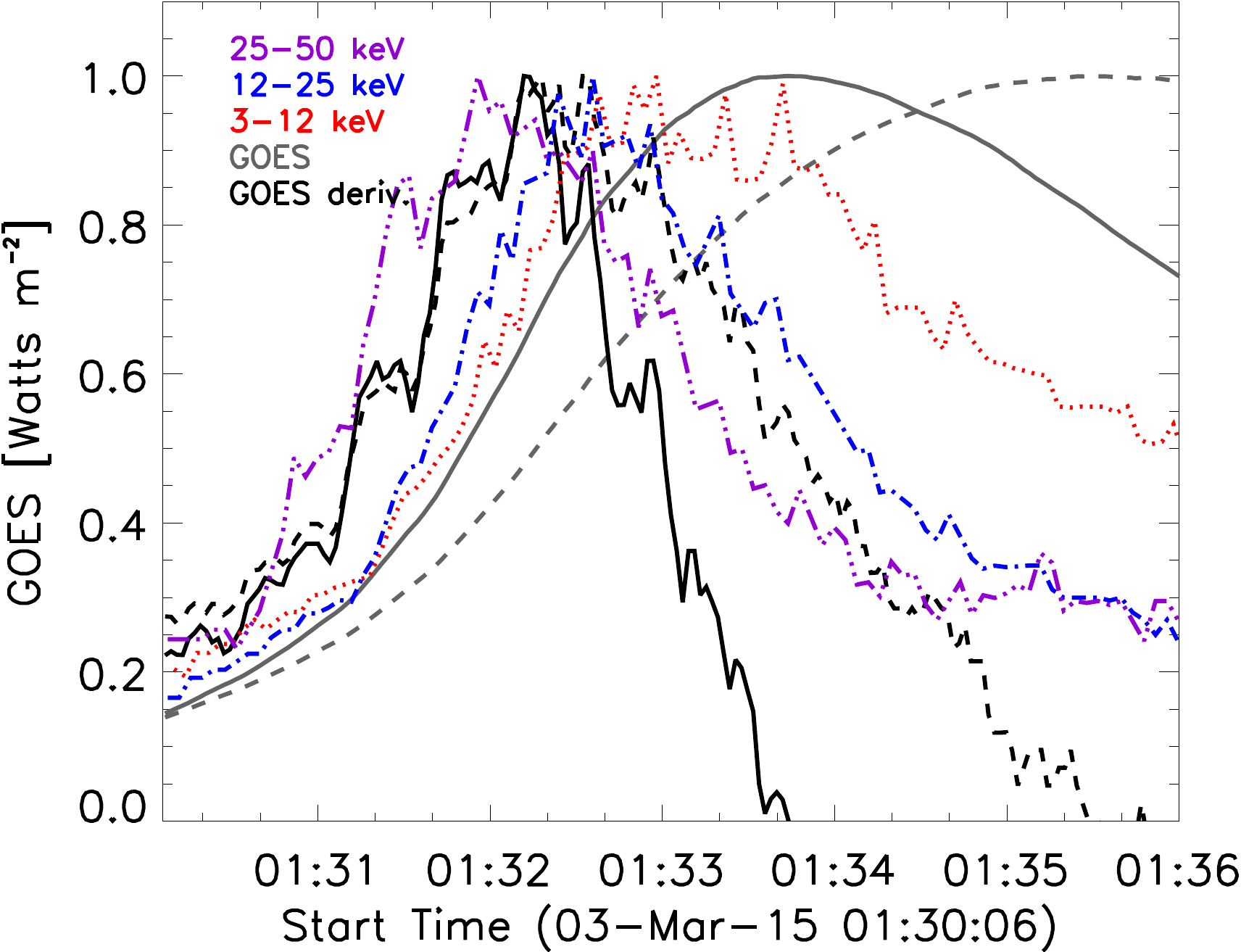}
\caption{Temporal evolution of the soft X-ray time derivative and the
  hard X-ray {\it RHESSI} count rates in three energy ranges (red,
  blue and purple). The {\it GOES} high energy ($0.5-4$~\AA) and low
  energy ($1-8$~\AA) fluxes are plotted as solid and dashed grey
  lines, while their derivatives are given by the respective black
  lines. All quantities are normalized to their maximum values in the
  time interval.}
\label{fig:lc_goes_rhessi}
\end{figure}

Figure~\ref{fig:lc_goes_rhessi} shows an example of the temporal
evolution of the soft and hard X-ray flux in different {\it GOES} and
{\it RHESSI} energy channels. It can be seen that the two {\it RHESSI}
lowest energy channels are delayed with respect to the {\it GOES}
derivatives. The high energy channel at $25-50$~keV has a quick rise
to maximum and correlates well with both {\it GOES} derivatives during
the rise phase. Later, during the decay phase, the lower energy soft
X-rays decay slower, implying a longer cooling timescale. A
cross-correlation analysis showed a correlation coefficient of 0.95
for this high-energy \emph{RHESSI} channel and the
$0.5-4$~\AA~\emph{GOES} derivative and no substantial lag. The
correlation of the low-energy \emph{GOES} channel with the RHESSI
$25-50$~keV band is slightly worse (0.80) but equally good when
comparing with the {\it RHESSI} $12-25$~keV light curve. This
represents a clear example of a strong correlation in our study.

\begin{figure*}[ht]
\centering
\includegraphics[width=0.45\textwidth]{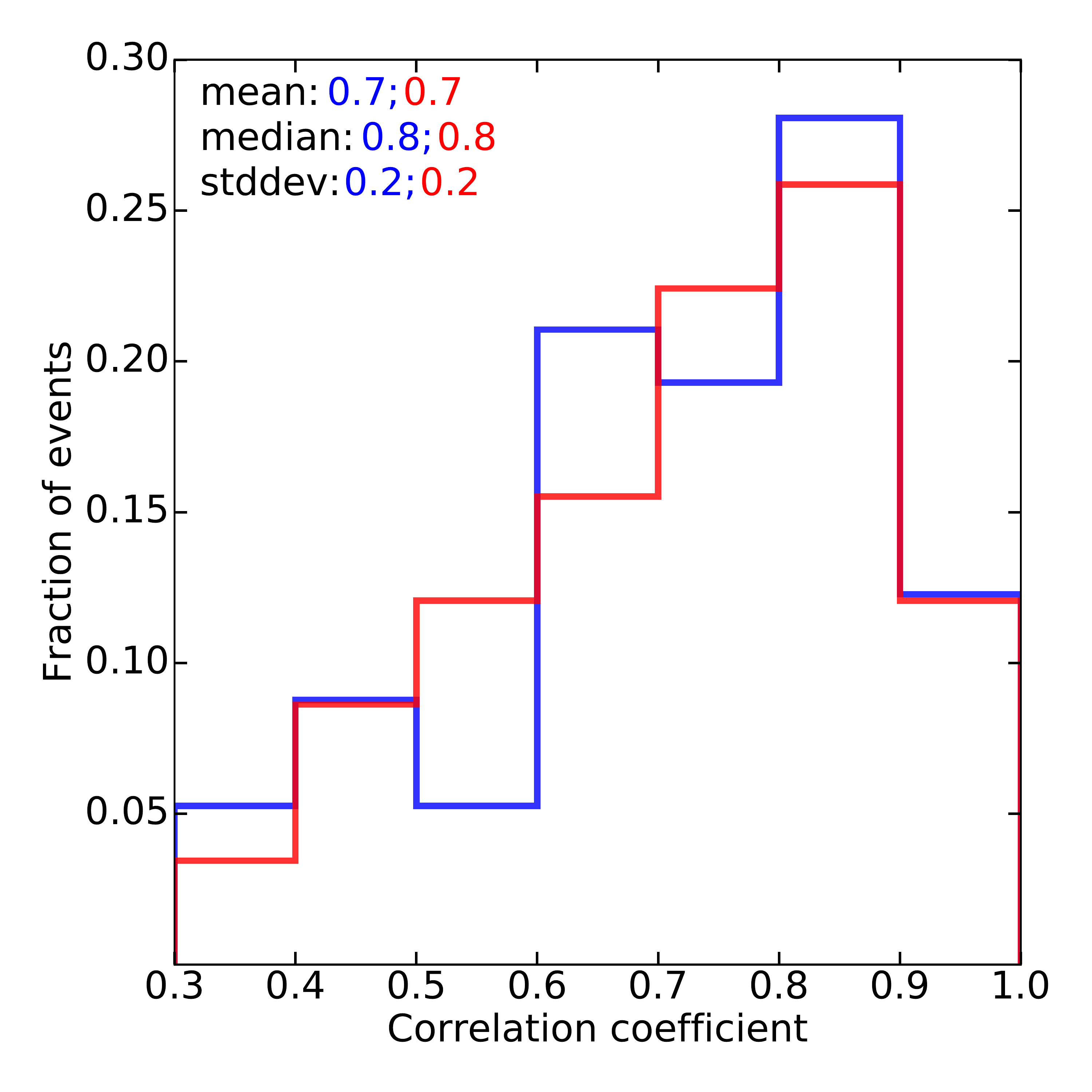}
\includegraphics[width=0.442\textwidth]{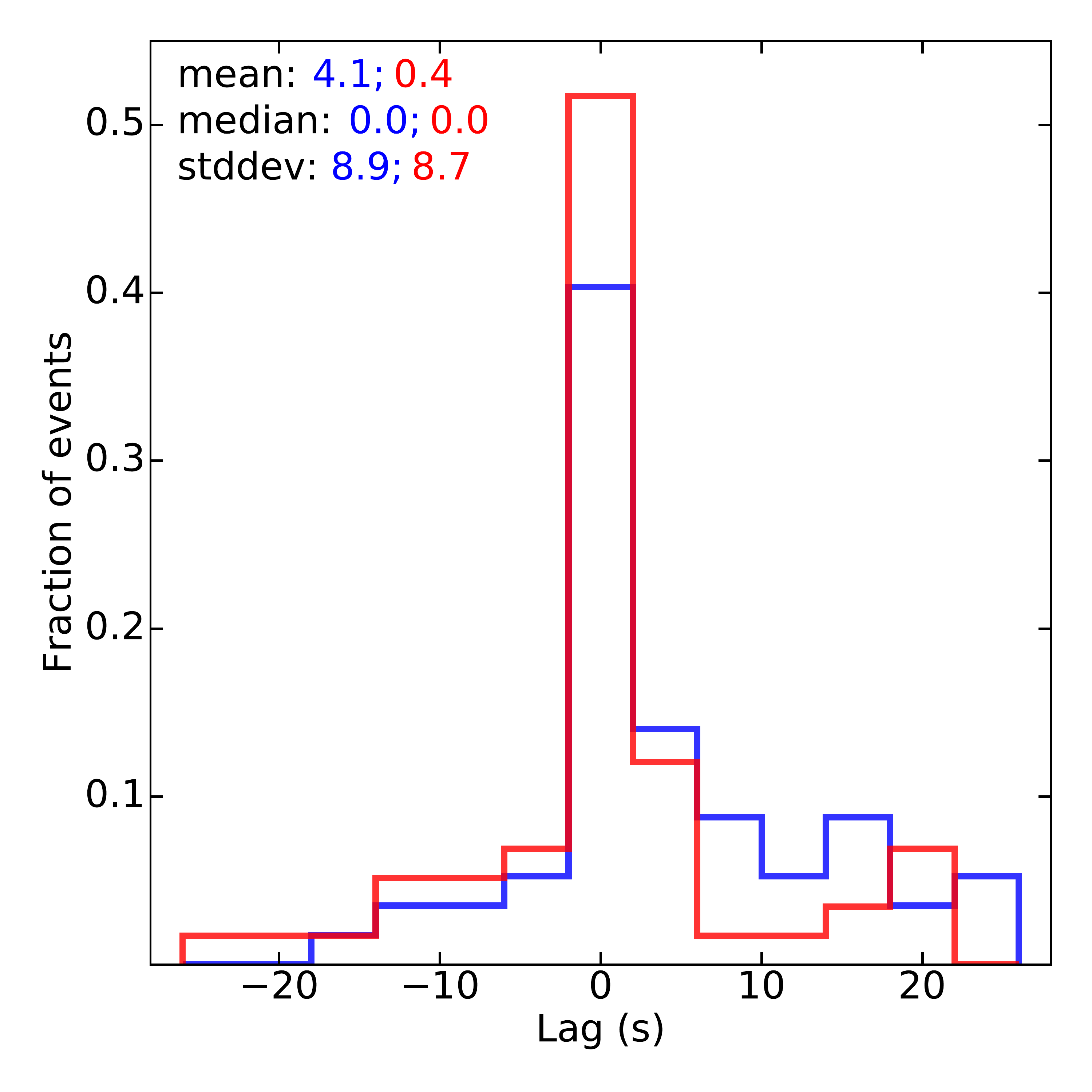}
\caption{Correlation coefficients (left) and temporal lags (in s,
  right) for our complete ensemble of partially occulted flares as
  calculated from the light curve cross-correlation analysis of the
  \emph{GOES} soft X-ray time derivative (low channel: red; high
  channel: blue) and the \rhessi hard X-ray flux. Positive lags imply
  an earlier rise in soft X-rays compared to hard X-rays.}
\label{fig:hist_corr_rhessi_goes}
\end{figure*}
By discarding all the thermal events (cf. Section~\ref{Sect:spectra})
and those with incomplete {\it GOES} or \emph{RHESSI} light-curve
coverage, 57 events remained for the {\it GOES} correlation analysis
in this study (the discarded values are labeled with a dash in
Tables~\ref{occulted_list} and
\ref{KL_list}). Figure~\ref{fig:hist_corr_rhessi_goes} presents a
histogram of the correlation and lag between the best-fitting high
energy \emph{RHESSI} channel and the {\it GOES} soft X-ray time
derivatives. Many flares show a good correlation and a small number of
lags has a tendency towards positive values meaning that the rise in
the derivative of the high energy channel soft X-rays occurs earlier
than the hard X-ray emission. Most of the flares with strong
correlations do not show a significant lag.

\subsection{Imaging}
\label{Sect:image} 
\emph{RHESSI}'s unique imaging capability allows a detailed study of
the spatial structure of the hard X-ray emission. We consider 20
seconds around the first peak of the flare in the highest energy range
with increased count rate to be our interval of interest (see column~3
of Tables~\ref{occulted_list} and \ref{KL_list}), avoiding attenuator
changes when needed. For every flare in our list, we created images in
a low energy (typically $\sim$ 6-14~keV) and high energy (typically
$>20$~keV) range, using the CLEAN algorithm \citep{Hurford-etal-2002}
and a combination of detectors suitable for imaging in that time
interval (usually a subset of detectors 3-8). This avoids detectors
not properly segmented at a given time. Some flares with no clear
high-energy signal (typically lower than 22 keV) did not allow for
such analysis. They were discarded from this part of the statistical
study. These `thermal' flares, as shown in the tables, have only
temperature values as derived from a purely thermal fit. When
selecting the range for the high-energy component, we carefully
checked that the break energy as inferred from the broken power-law
spectral analysis (cf. Section~\ref{Sect:spectra}) is at least 4~keV
(or 4 energy bins) lower than the lower boundary of our energy
interval.

Apart from confirming that there is no visible footpoint emission for
a particular flare during this time, these images allow to estimate
the \emph{radial} separation between the thermal and non-thermal
emission. We determined the distance $d_{max}$ between the maxima and
distance $d_{com}$ between centers of mass of the low and high energy
images. Positive values indicate that the non-thermal source is
located farther away from the limb than the thermal component. The
resulting values for $d_{max}$ are reported in
Tables~\ref{occulted_list} and \ref{KL_list}. The values for $d_{com}$
are generally very similar.

\begin{figure*}[ht]
\centering
\includegraphics[width=0.47\textwidth]{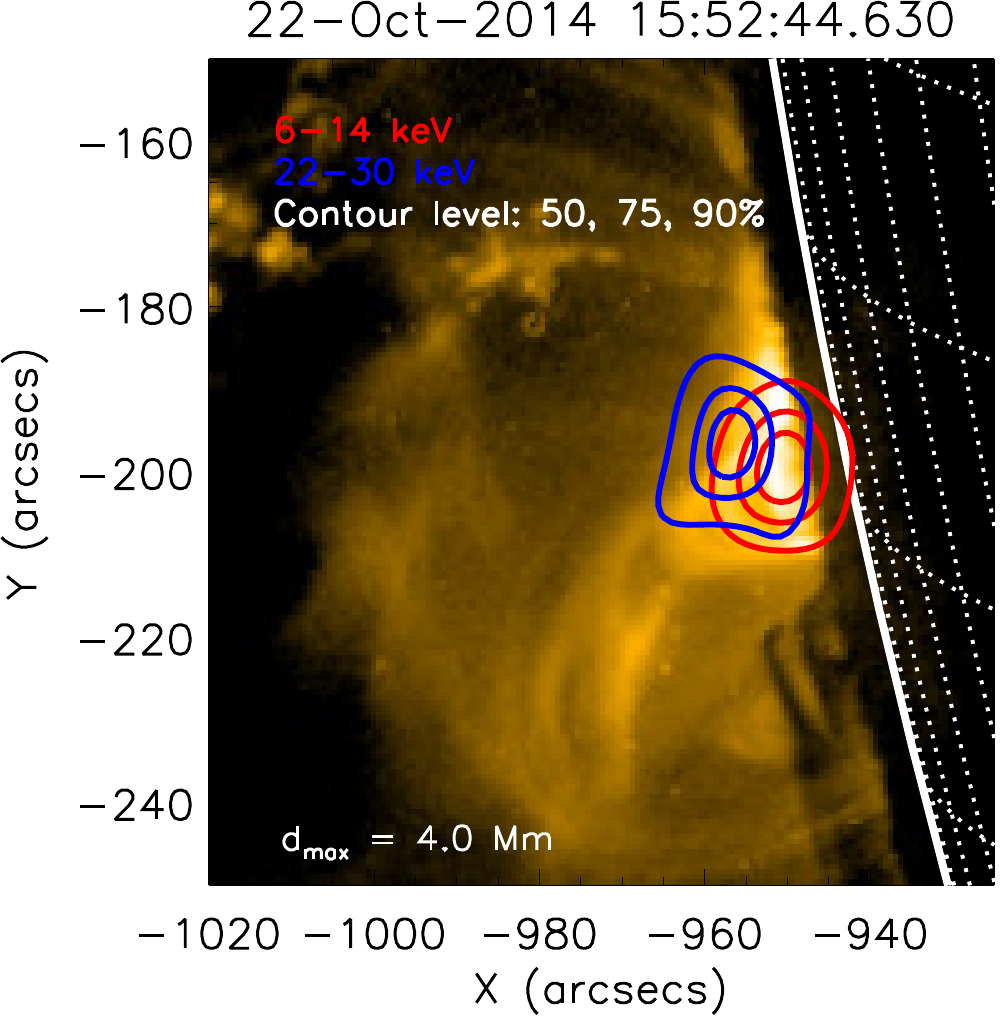}\hspace{0.2in}
\includegraphics[width=0.47\textwidth]{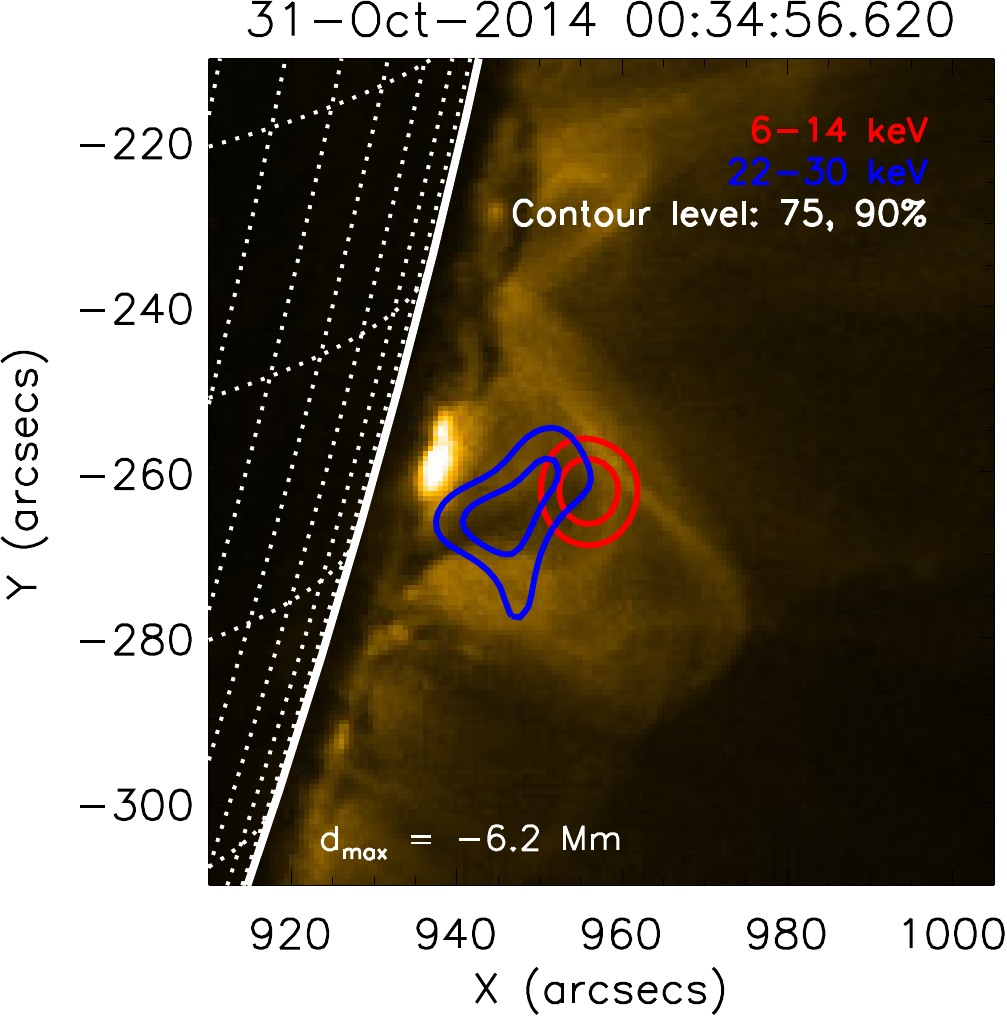}
\caption{Background AIA~131~\AA\ emission and \emph{RHESSI} X-ray
  (CLEAN algorithm) contours at 6-14~keV
  (red) and 22-30~keV (blue). Left: Coronal emission of the M1.4 class
  flare that occurred on October 22, 2014 with high energies at larger
  radial distance than low energies.  Right: Emission of the C8.2
  class flare from October 31, 2014 having inverted radial positions.}
\label{fig:clean_im}
\end{figure*}
Figure~\ref{fig:clean_im} (left) shows an image of the October 22,
2014 M1.4 class flare as an example. We find a positive radial
separation of about 4 Mm between the two energy maxima, meaning that
the non-thermal source is at higher altitude. The coronal emission at
131~\AA\ shows multiple bright loops. The higher energy, mostly
non-thermal X-ray emission is near the top of the coronal loops. Other
AIA wavelength don't show the loops as clearly, indicating that they
are hot with temparatures of about 10~MK. This radial ordering of low
and high energy emission is expected from standard flare scenarios,
since the non-thermal particles are presumably produce close to the
reconnection region above the thermal loop top
\citep[e.g.][]{Krucker-etal-2008}.

However, we found for a few flares in our sample that the high-energy
emission is centered closer to the limb than the lower energies. The
right panel of Figure~\ref{fig:clean_im} gives an example. The C8.2
class flare from October 31, 2014 has an extended high-energy emission
region below a relatively compact thermal source, which is about 6 Mm
higher in the corona. In this particular case, the high-energy
emission coincides with a dark structure in the 131 \AA\ AIA channel
(also clearly visible e.g. in 335 \AA). A possible explanation for the
emission there is thus that it acts as dense target above the
chromosphere for the non-thermal particles. Alternatively, the bright
EUV emission close to the limb could indicate that we only see the
above-the-looptop part of the X-ray emission
\citep[e.g.][]{Liu-etal-2013}, which would show such an inverted
ordering of low and high energies.\footnote{One should keep in mind
  that, as previously noted by \citet{Kuhar-etal-2016}, there is a
  spatial separation of $\approx$~2.5 arcsec between AIA and
  \emph{RHESSI}, most probably due to an error in the roll-angle
  calibration. Since we only use \rhessi data for the quantitative
  analysis, and the calculation of $d_{max}$ and $d_{com}$ is not
  affected, this is of no concern for our study, apart from the
  overlay imaging.}

\begin{figure}[t]
\centering
\includegraphics[width=0.45\textwidth]{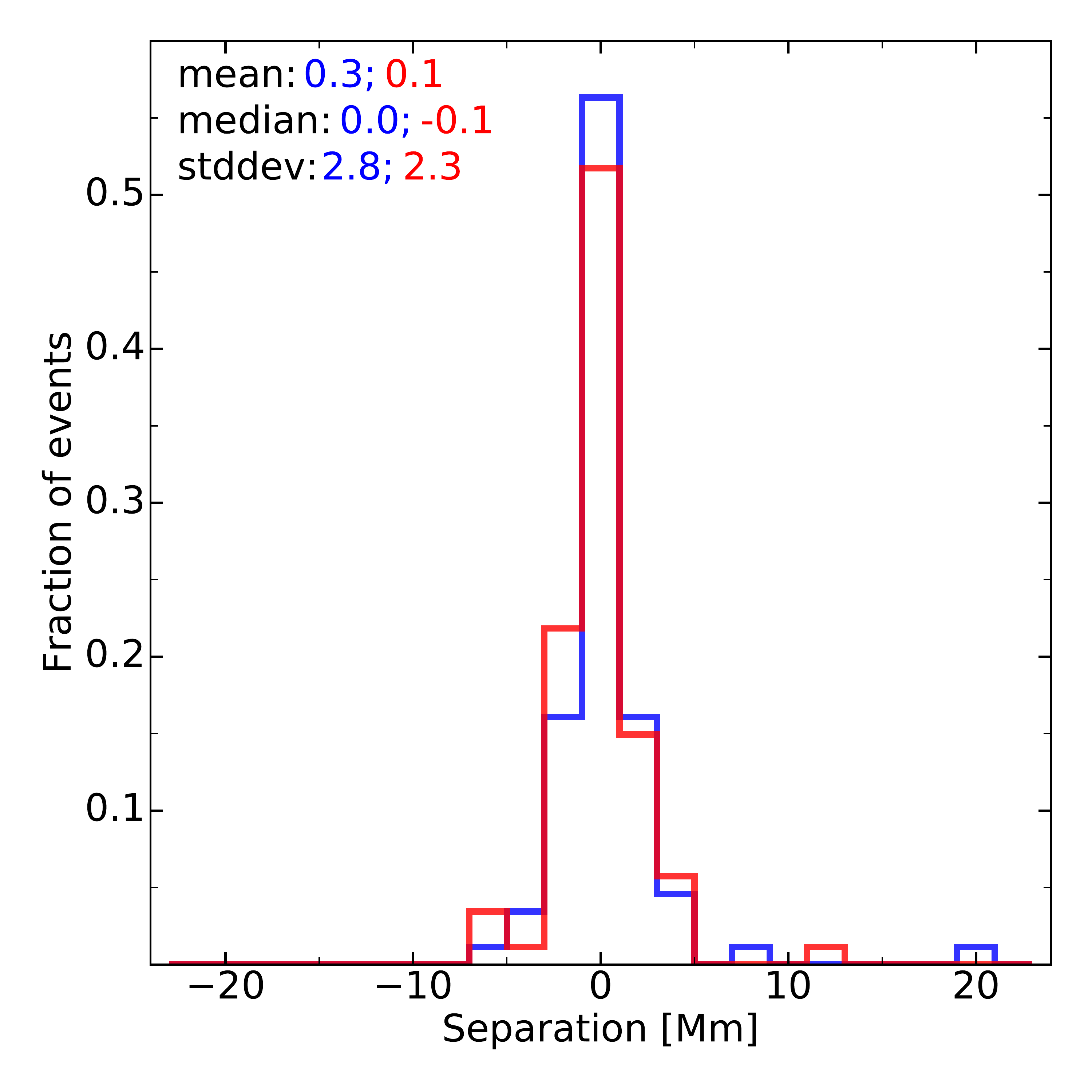}
\caption{Frequency distribution of the separation between the low and
  high energy X-ray sources as quantified by their emission maximum
  ($d_{max}$, blue) and center of mass ($d_{com}$, red), resulting
  from the imaging analysis. Positive values indicate that the
  non-thermal source is located farther away from the limb than the
  thermal component.}
\label{fig:hist_separation}
\end{figure}
Figure~\ref{fig:hist_separation} gives a histogram of separation
estimates $d_{max}$ and $d_{com}$ between low and high energies. The
two estimates do not differ significantly from each other. We find no
clear tendency towards positive or negative separations between the
low and high energy sources. The mean of $d_{max}$ is 0.3~Mm,
indicating a possible trend that the higher energy emission might
radially be farther out in the corona, but this value is still
consistent with no separation.

\subsection{STEREO Analysis: Height and GOES-class}
\label{Sect:stereo}
The twin \emph{STEREO}-A and B spacecraft allowed us to confirm for
many of the flares in our cycle~24 sample that the associated active
regions and footpoints were indeed located behind the limb. We were
also able to estimate the heights of the X-ray sources and
the true (un-occulted) soft X-ray magnitudes of the flares. However,
depending on the {\it STEREO} positions, and the quality and cadence
of their data, these estimates were not always obtainable,
particularly from October 2014 to November 2015, when both
\emph{STEREO} spacecraft were on the opposite side of the Sun near the
Sun-Earth line and had limited or no telemetry.

\begin{figure}[t]
\centering
\includegraphics[width=0.49\textwidth]{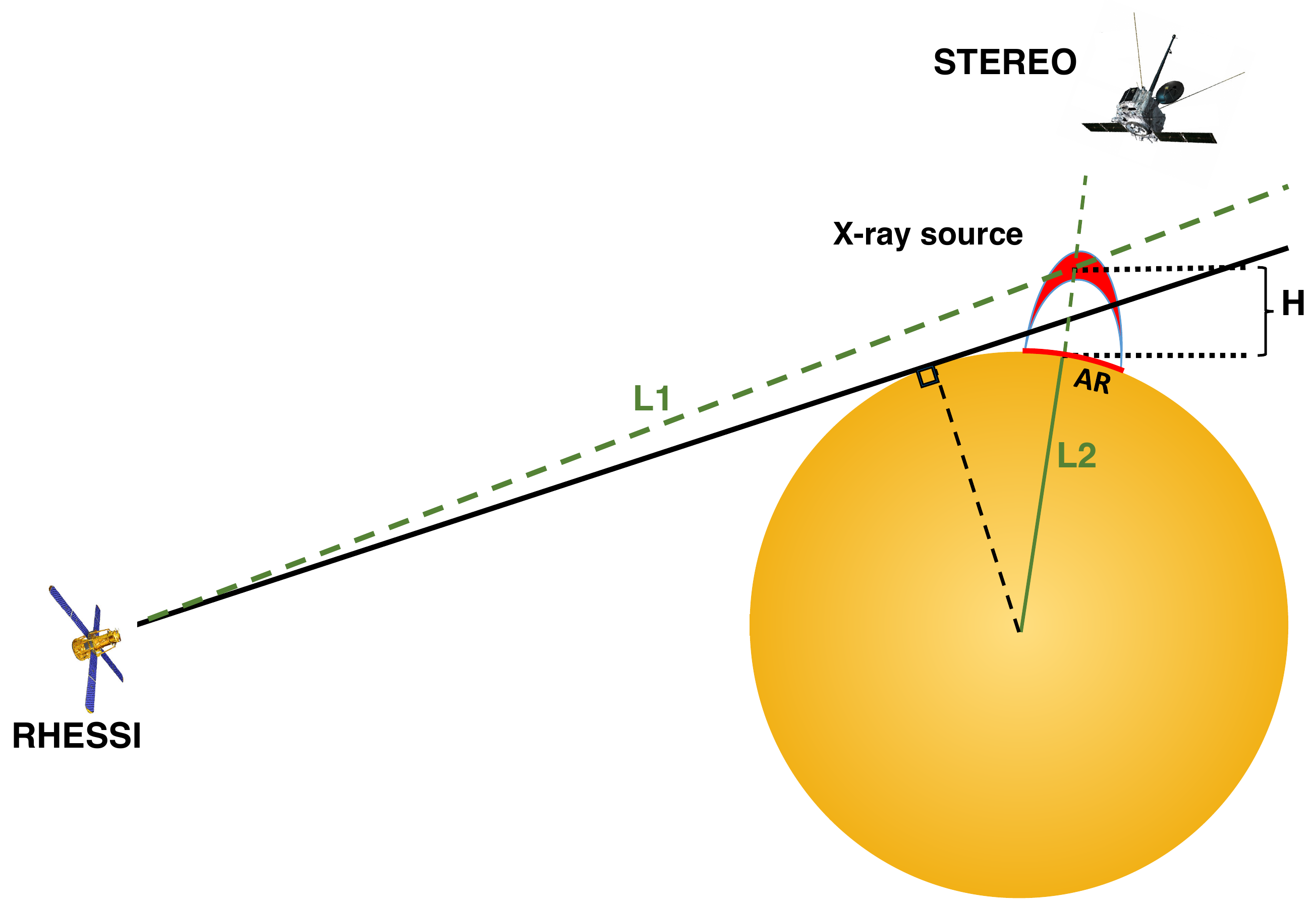}
\caption{Two-dimensional sketch to illustrate the geometry of view
  angles and the calculation of the flare height $H$ with combined
  \emph{RHESSI} and \emph{STEREO} information.}
\label{fig:stereo-cartoon}
\end{figure}

By combining the line of sights of the emissions seen by {\it RHESSI}
and {\it STEREO} into a 3D structure, we estimated the height of the
coronal emission. The geometry is illustrated in
Figure~\ref{fig:stereo-cartoon}.  The line-of-sight from Earth towards
the source (`L1') and the radial (`L2') from the center of the Sun
through the brightest point of the active region (AR), as selected
from \emph{STEREO} observations, do not necessarily intersect in 3D
space (the X-ray source may not be directly above the AR). To estimate
a source altitude above the photosphere, a vector was drawn from the
AR to the midpoint of the shortest possible line segment connecting
both L1 and L2. The projection of that vector on the local vertical
gives the estimated heights $H$ of the coronal emission. These
  values are given in column 8 of Table~\ref{occulted_list}, with a
  mean of 14 Mm and a median height of 11.3 Mm, consistent with a
  typical loop size.

The temporal evolution of the {\it STEREO}~195~\AA~emission allowed us
to extrapolate to other wavelengths and estimate the soft X-ray
magnitude of the flare, as if it would be an on-disk event.  This was
done using the empirical relation between the peak {\it
  STEREO}~195~\AA\ flux and the \emph{GOES} 1\,--\,8~\AA\ soft X-ray
flux \citep[][Eq.~(1) and their Figure~7]{Nitta-etal-2013}:
\begin{equation}
\label{Nitta-eq}
  F_{\rm GOES} = 1.39 \times 10^{-11} F_{\rm EUVI
    (195)},
\end{equation}
where $F_{\rm GOES}$ is the \emph{GOES} 1\,--\,8~\AA\ channel flux in
units of [W~m$^{-2}$] and $F_{\rm EUVI (195)}$ is the pre-event
background subtracted \emph{STEREO} 195~\AA\ flux in
[DN~s$^{-1}$].

The resulting \emph{GOES} class estimates are reported in column~5 of
Table~\ref{occulted_list}, along with the actual \emph{GOES} class
observed from Earth in column~4. In general, these estimates show that
the latter often significantly underestimate the true magnitude of the
flare. However, we note that there are a few outliers from this
general trend.  That is, the estimated \emph{GOES} class can be lower
in some instances than the observed class from Earth. There are at
least two possible causes of this discrepancy. First,
equation~\ref{Nitta-eq} is an empirical relation that has certain
ranges of uncertainties, which are within a factor of three for flares
$>$M4 class and an order of magnitude for less intense flares.
Second, in case of low-cadence ($\ge$10~minutes) observations,
\emph{STEREO} can miss the true EUV peak and thus underestimate the
flare class.

\subsection{Spectral analysis}
\label{Sect:spectra}
Two kinds of spectral analysis were performed for every
flare in our list, followed by detailed checks of the goodness-of-the
fit and re-analysis when necessary. All fits were done with the
standard Object Spectral Executive (OSPEX) software package
\citep[e.g.][]{Schwartz-etal-2002}. The fitting time
interval is the same 20s around the first non-thermal peak as
described for imaging. By using an initially automated procedure, we
have a better comparability of fitting results between different
flares.

(1) The first fitting model is a fit of the observed photon spectrum
by a thermal plus broken power-law model (hereafter, th-bpow), similar
to that used in KL2008, which has five free parameters: the emission
measure, $EM_{th}$, and temperature, $T_{th}$, of the thermal
component, the normalization, $A_\gamma$, the break energy
$E_{break}$, and the spectral index, $\gamma$, above the break of the
power-law component,
$I_\gamma(\epsilon) = A_\gamma \epsilon^{-\gamma}$.  The index below
$E_{break}$ is fixed to 1.5 \citep{Holman-etal-2003} and the relative
abundances in the thermal component are kept at 1.
  
(2) The second method fits the observed photon spectrum by
bremsstrahlung emission arising from a kappa spectral model for the
flux of (non-relativistic) accelerated electrons
\citep[th-kappa,][]{Kasparova-Karlicky-2009}:
\begin{equation}
  F_\kappa(E) = A_\kappa \frac{{E}}{\sqrt{(k_BT_\kappa)^3}}
  \left( 1 + \frac{E}{(\kappa - 1.5)k_BT_\kappa}\right)^{-(\kappa+1)}\,.
\end{equation}

This model, with three parameters, is a generalization of a
non-relativistic Maxwellian distribution with an enhanced non-thermal
tail approaching a power law with index $\kappa$ at high energies.
However, its thermal component is often not strong enough, especially
for low values of the index $\kappa$, to reproduce the prominent
thermal component of solar flares at low energies. As a result we had
to add an additional thermal component with two additional free
parameters, emission measure and temperature,
$EM^\kappa, T_{th}^\kappa$, again giving five free fitting parameters
\citep[see][for detailed case studies including an additional thermal
component for coronal sources]{Oka-etal-2013,Oka-etal-2015}.

An explanation for the necessity of an additional thermal component
could be that the emission of the chromospherically evaporated plasma
is superimposed onto the in-situ heated component of the kappa
distribution in the corona. Imaging spectroscopy can separate
different thermal (and non-thermal) parts of the emission
\citep{Oka-etal-2015}, but since most thermal and non-thermal sources
are co-spatial (cf.\ Section~\ref{Sect:image}), in practice this is
usually not possible. \citet{Battaglia-etal-2015} recently improved
the estimation of thermal components by combining emission measures
from \emph{RHESSI} and AIA. This approach may enable further insights
into the thermal part of the electron population of coronal sources in
future studies.

A key feature of our study is that we fit all available detector
spectra separately and combine the resulting parameters of the fit
into average quantities. This approach, as detailed in
\citet{Liu-etal-2008} and \citet{Milligan-Dennis-2009}, takes advantage
of the fact that each detector provides an independent measurement of
the X-ray spectrum and avoids smearing in energy of slightly different
detector responses. We individually discarded certain detectors for
every flare that did not perform properly or showed otherwise strong
deviations from the average results.

\begin{figure*}[t]
\centering
\includegraphics[width=0.40\textwidth]{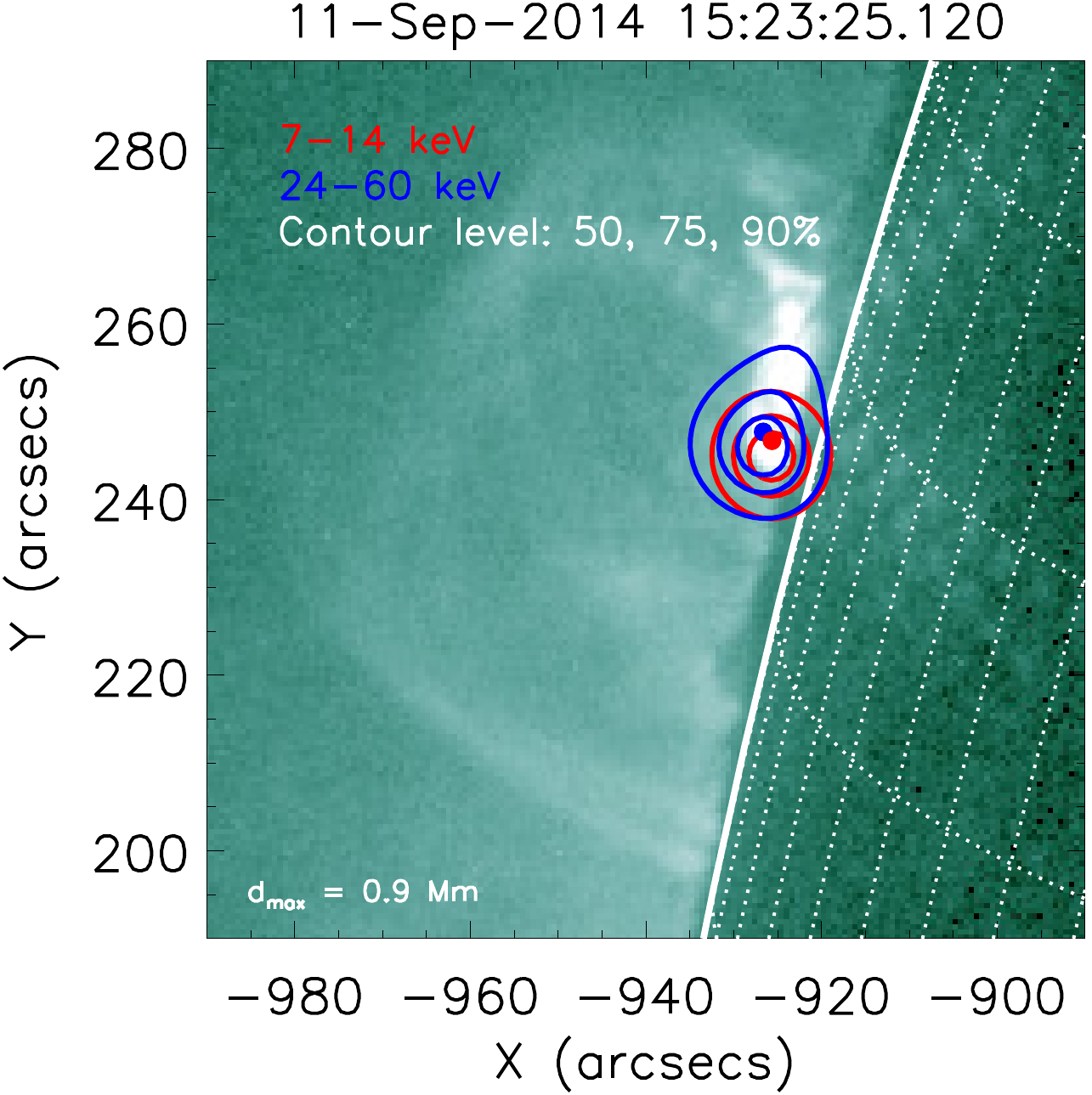} \hspace{0.2in}
\includegraphics[width=0.45\textwidth]{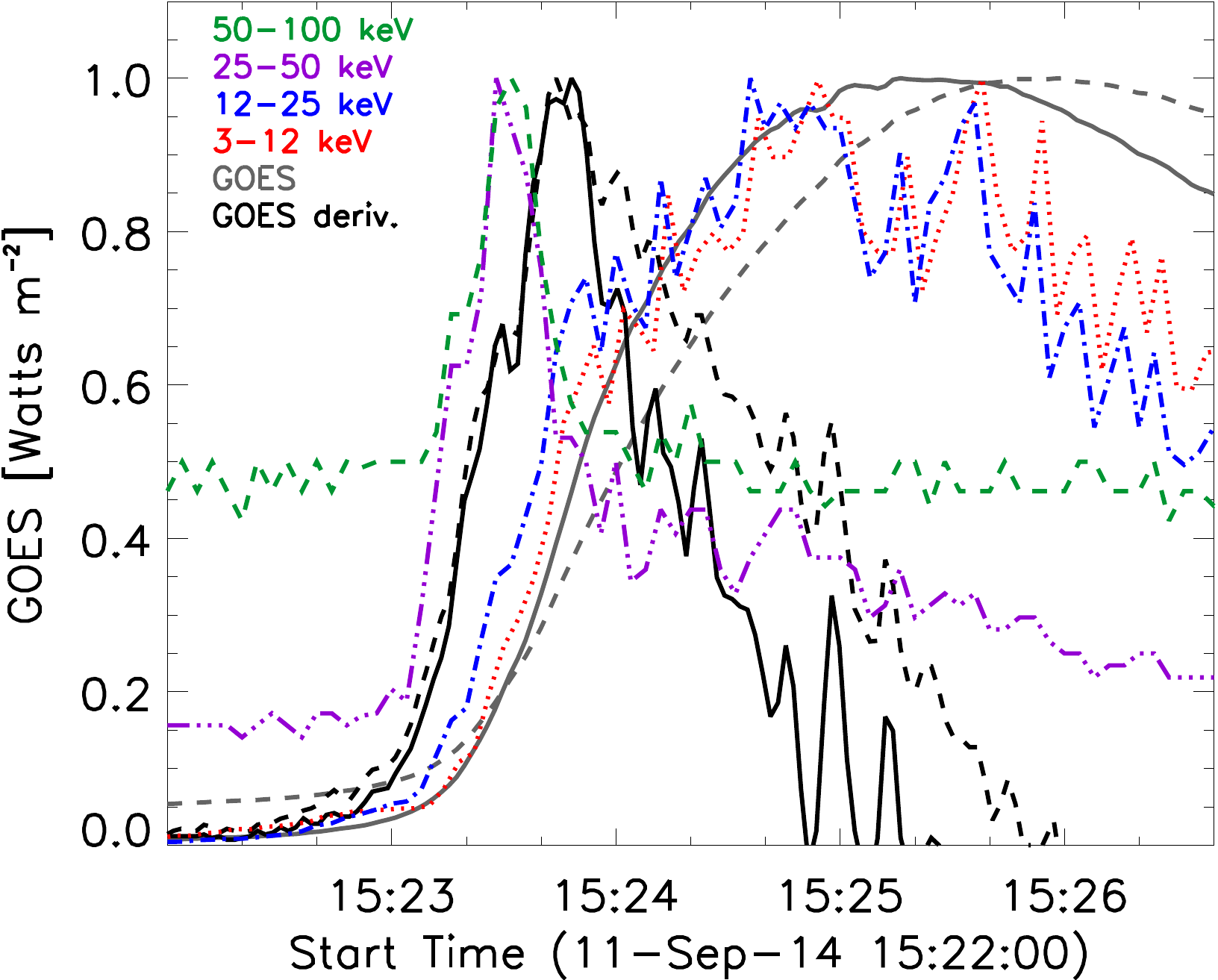}\\
\includegraphics[width=0.45\textwidth]{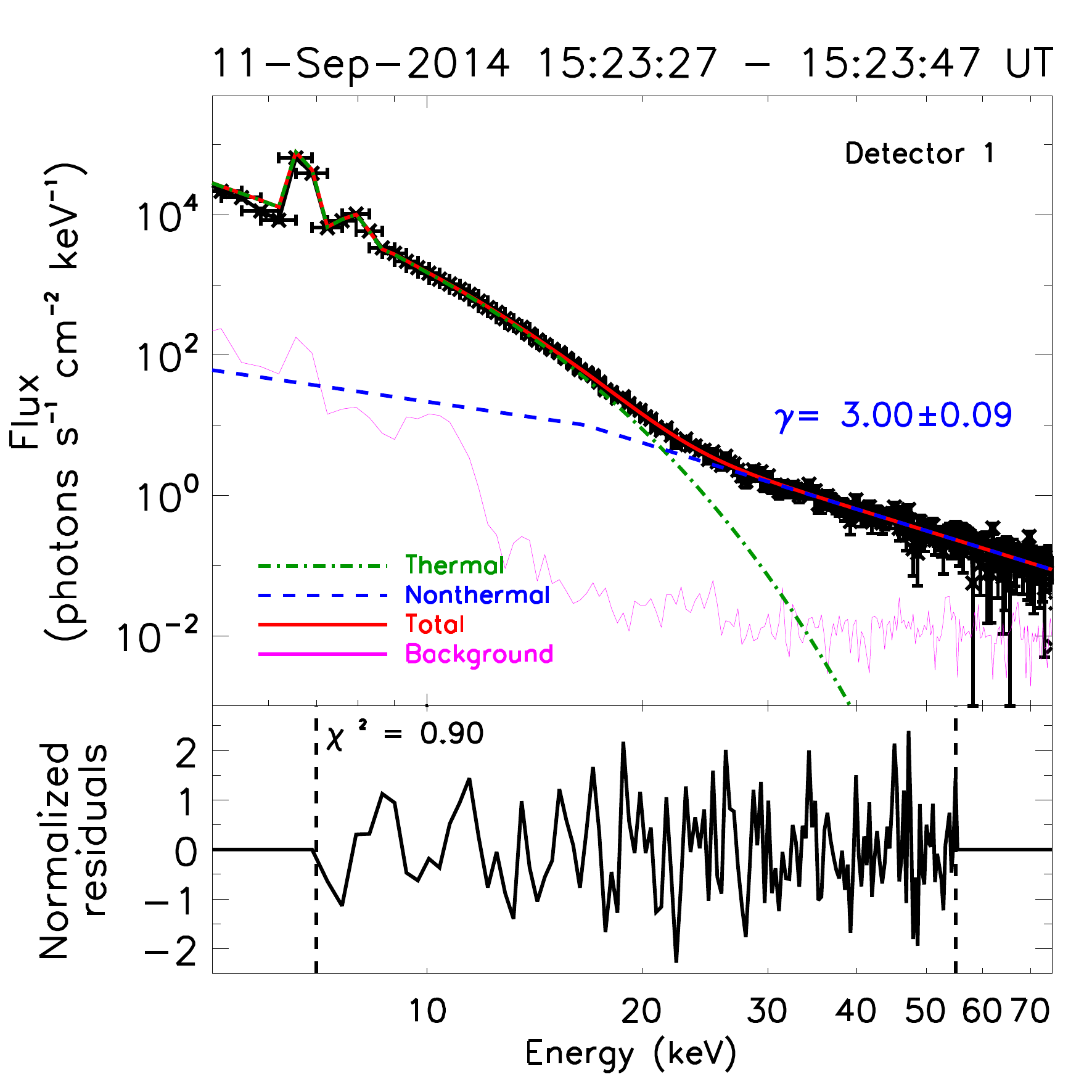}
\includegraphics[width=0.45\textwidth]{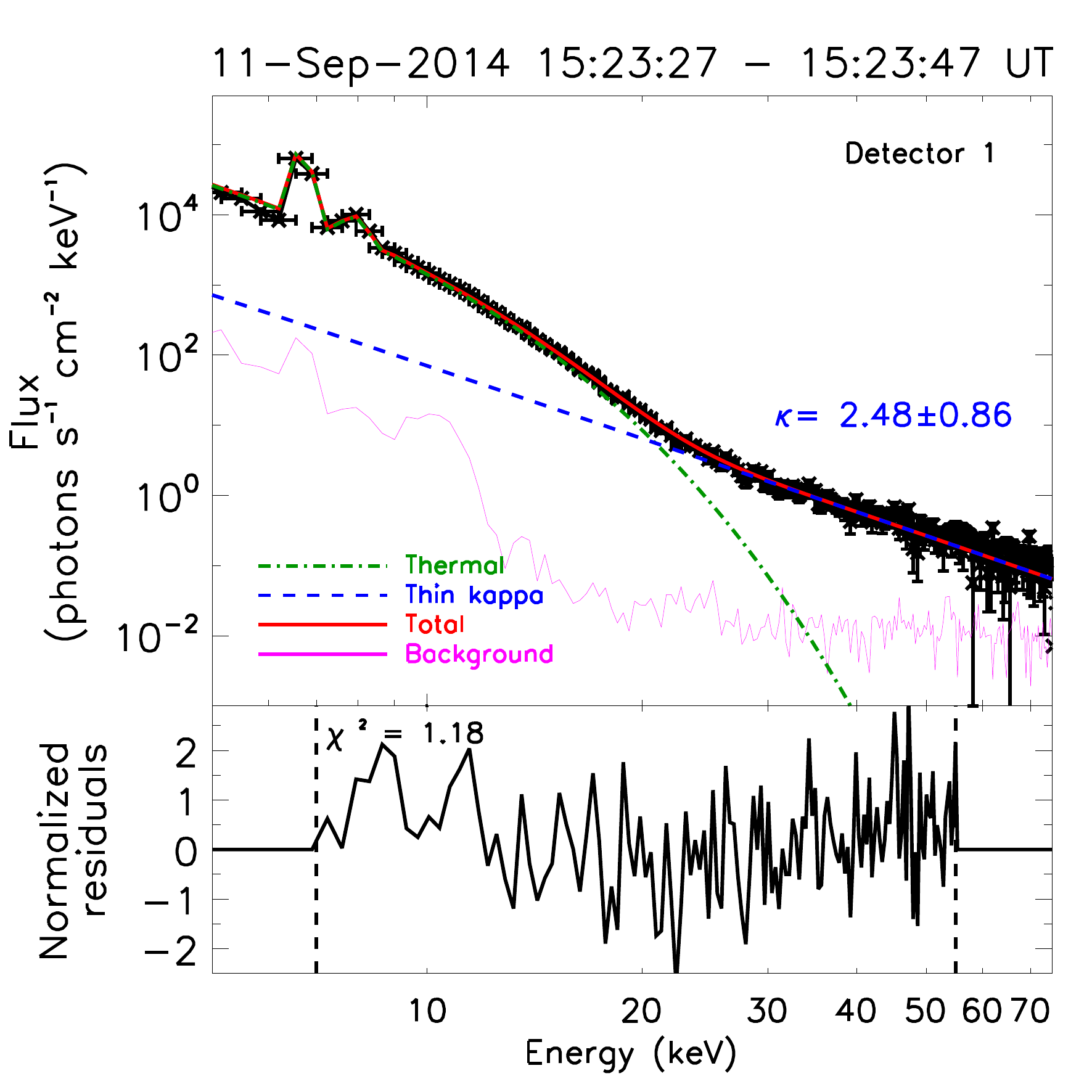}
\caption{Top left: Coronal emission of the September 11, 2014 M2.1
  flare as observed by AIA at 94~\AA. The contours correspond to the
  X-ray emission at low (7-14~keV, red) and high (24-60~keV, blue)
  energies, using the CLEAN algorithm of \citet{Hurford-etal-2002}
  integrated over 20s around the peak in detectors 3-8. Top right:
  Light curves of {\it RHESSI} count rates at four energy ranges
  (red, blue, purple and green), {\it GOES} high energy flux (0.4
  to 5~\AA, grey, dash-dotted) and its time derivative (black,
  solid). The counts in the two high energy channels are multiplied by
  20 and 15, respectively, to make them comparable in
  magnitude. Bottom: Photon spectra as observed by {\it{RHESSI}}
  detector~1 (15:23:27-15:23:47~UT). The spectrum has been fitted to a
  thermal component plus a broken power law (left) and a thermal
  component plus a thin-target kappa distribution function (right).}
\label{fig:spectra_fit}
\end{figure*}
Figure~\ref{fig:spectra_fit} shows example fit results for both
fitting approaches applied to the M2.1 class flare occurred on
September 11, 2014, together with the corresponding light curve and
imaging analysis. There is no significant spatial separation between
thermal and non-thermal coronal emission in this flare and the light
curve shows a quick onset of high-energy X-rays with a slightly
delayed response in the {\it GOES} derivative. The first smaller peak
in the {\it GOES} derivative is temporally related to the onset of the
highest energy (25-100 keV) X-rays detected by \emph{RHESSI}, while
the second, larger peak is associated with a peak at lower energies. The 6-12
keV \emph{RHESSI} emission aligns well with the temporal evolution of
the soft X-rays detected by {\it GOES}. Both spectral fits with a
broken power law and thin-target kappa function result in a low
$\chi^2$ value and a good fit over all energies as indicated by the
residuals. The high energy broken power-law spectral index $\gamma$
agrees with the expected electron $\kappa$ index within the estimated
standard deviation for a thin-target model (see also the discussion
below).

The results of our spectral analysis are compiled in
Tables~\ref{occulted_list} and \ref{KL_list}, with vertical solid
lines separating the two groups of fitting parameters. We note that
for some flares, one or both approaches did not converge
satisfactorily to a final set of parameters, in which case we left the
table empty for these values (`-'), and discarded them from the
statistical analysis.

``Thermal'' flares, without clearly distinguishable power-law
component at higher energies were fitted only to a pure thermal
component. The resulting temperature is reported as $T_{th}$ in both
tables instead of the thermal component of the broken power-law fit.

\begin{figure*}[t]
\centering
\includegraphics[width=0.4\textwidth]{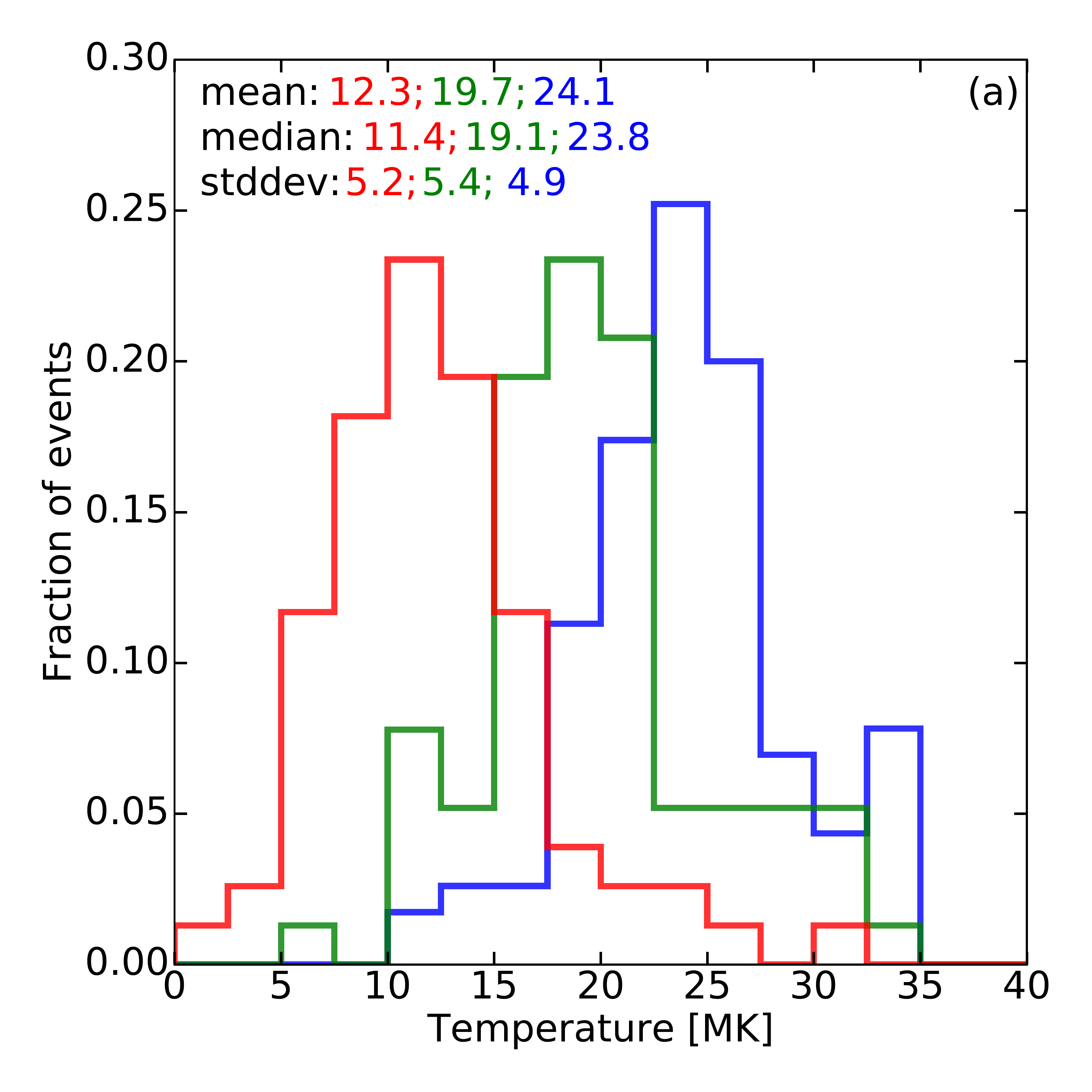}
\includegraphics[width=0.4\textwidth]{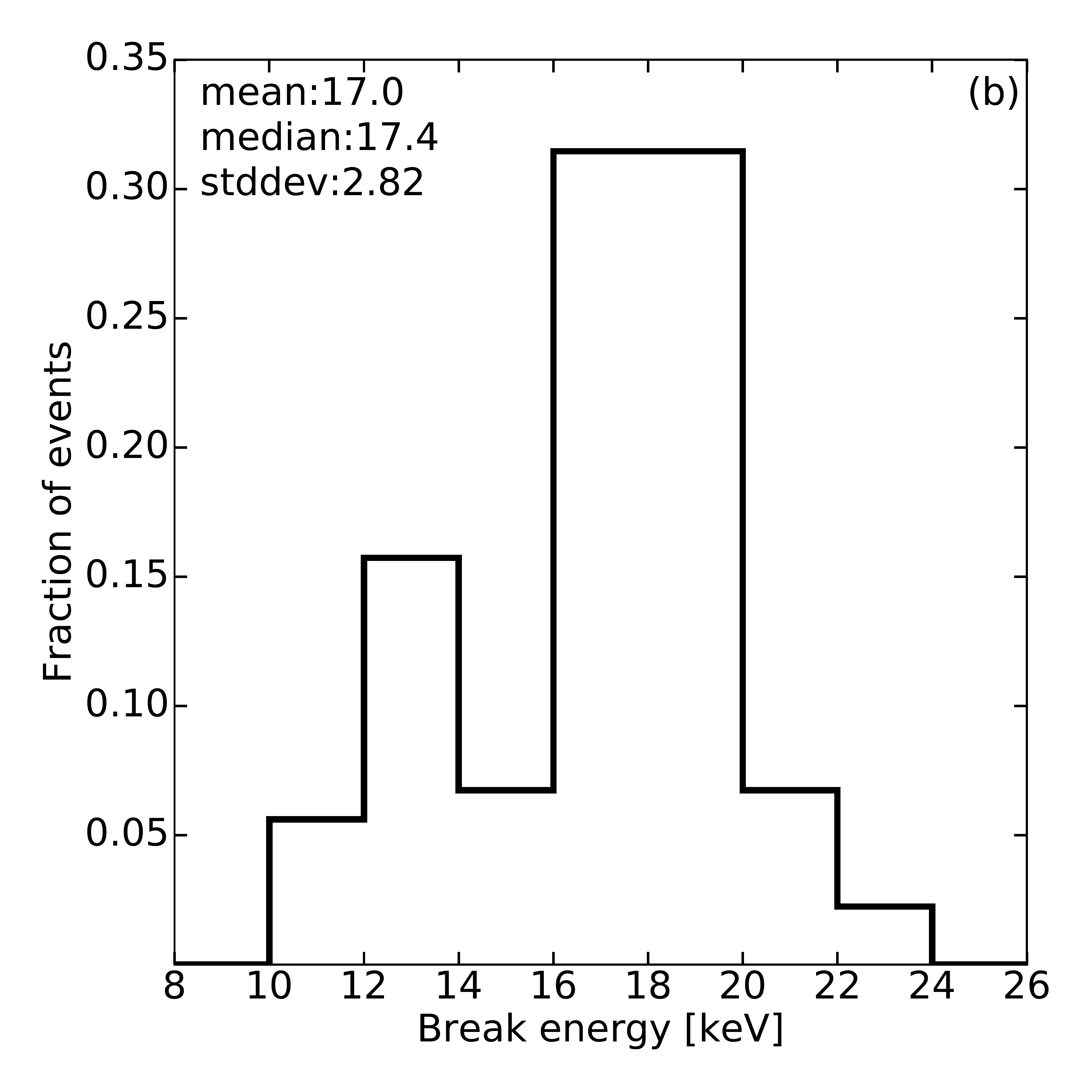}\\
\includegraphics[width=0.4\textwidth]{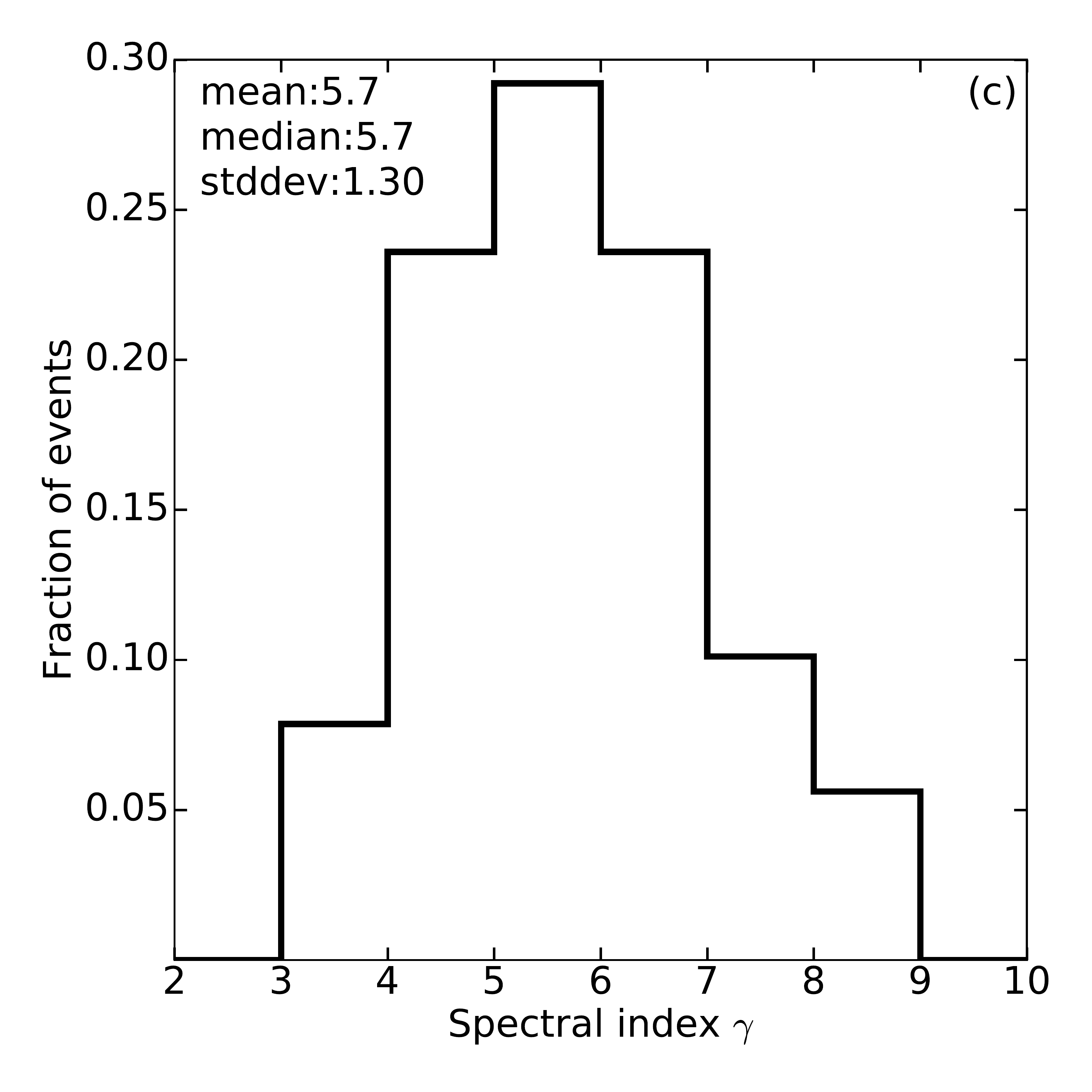}
\includegraphics[width=0.4\textwidth]{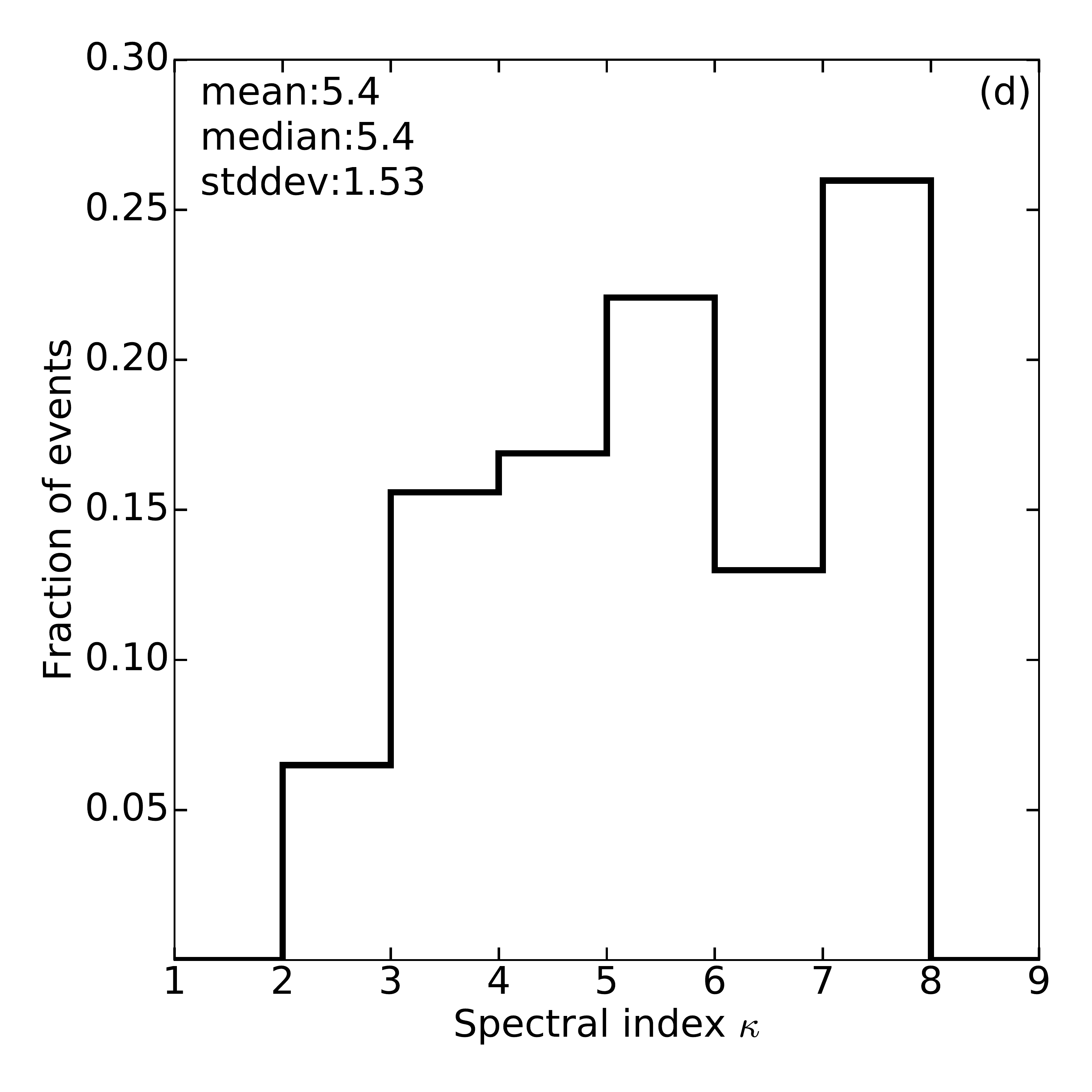}
\caption{Frequency distributions of fitting parameter results: (a) The
  three different temperatures, namely the thermal component
  temperature in the broken power-law fit, $T_{th}$ (blue), the
  thermal component temperature in the kappa fit, $T_{th}^{kappa}$
  (green), and the kappa-temperature $T_{\kappa}$; (b) Break energy of
  the broken power law $E_{break}$; (c) spectral index $\gamma$ above
  the break; (d) $\kappa$ values for the th-kappa model.}
\label{fig:histograms}
\end{figure*}
Figure~\ref{fig:histograms} gives an overview of the statistical
properties of the fitting results for our flare sample, combining both
solar cycles. The average temperatures are generally ordered from low
to high in $T_{\kappa}$, $T_{th}^{\kappa}$, and $T_{th}$.  This is
most likely due to the fact that the kappa distribution itself has
already a thermal contribution. The results for break energies and
power-law spectra indexes are in general agreement with the previous
results from KL2008, with a tendency to lower break energies in our
study. The spectral index $\kappa$ has a broader distribution. The
mean values are $\left\langle\gamma\right\rangle = 5.7$, similar to
the previously reported value in KL2008 of 5.4, and
$\left\langle\kappa\right\rangle = 5.4$. These are softer than what is
found for the high energy index of disk flares, which contain the
footpoint emission with harder spectrum
\citep[e.g.][]{McTiernan-Petrosian-1991,Saint-Hilaire-etal-2008,Warmuth-Mann-2016a,Warmuth-Mann-2016b}.

\section{Discussion}
\label{Sect:discussion}
We now discuss aspects of our flare sample that offer insights into
the coronal X-ray source structure and associated energetic electron
properties.

As previously mentioned in Section~\ref{Sect:stereo}, we verified that
the emission visible from the \rhessi field-of-view had no footpoint
contamination according to \emph{STEREO}, but this approach is also
influenced by the location of maximum EUV emission selected in the
active region. On the other hand, there are only 36~flares with viable
height information from \emph{STEREO}, leaving the decision on
possible chromospheric contamination to the available \rhessi and AIA
images, for which we verified that there was no on-disk signatures of
footpoints. Thus in what follows we will assume that we are dealing
with loop-top emissions in all the flares in the two samples.

It should be also noted that there are differences between the results
reported in this study and in KL2008 for the same set of 55 flares
(see Table~\ref{KL_list}). This is due to the combination of several
aspects: Our spectral fit approach is partially automated and the
initial fitting values and constraints of the variables used for the
spectral fits have not been changed unless it appeared
necessary. Moreover, the background subtraction and the exact choice
of the 20~second fitting interval, aiming for the first peak of the
fast time variation component, can influence the results further.

\subsection{Thermal vs. non-thermal energy flux}
As evident from Tables~\ref{occulted_list} and \ref{KL_list}, a
majority of flares have a thermal and a non-thermal component. In
general, weaker {\it GOES} class flares show a tendency to have only a
thermal component and in our two samples, about 20\% of flares show no
clear non-thermal part.  However, most of those are in the new
selection from solar cycle 24. Less than 8\% of cycle 23 but nearly
40\% of cycle 24 are in this category.  Part of this difference could
be related to the reduced efficiency of the \emph{RHESSI} detectors, in
particular during the later part of 2015 before the anneal procedure
in early 2016. On the other hand, our analysis of the time dependence
of the spectral fitting parameters shows no significant differences
between cycle 23 and 24.

\begin{figure}[t]
\centering
\includegraphics[width=0.5\textwidth]{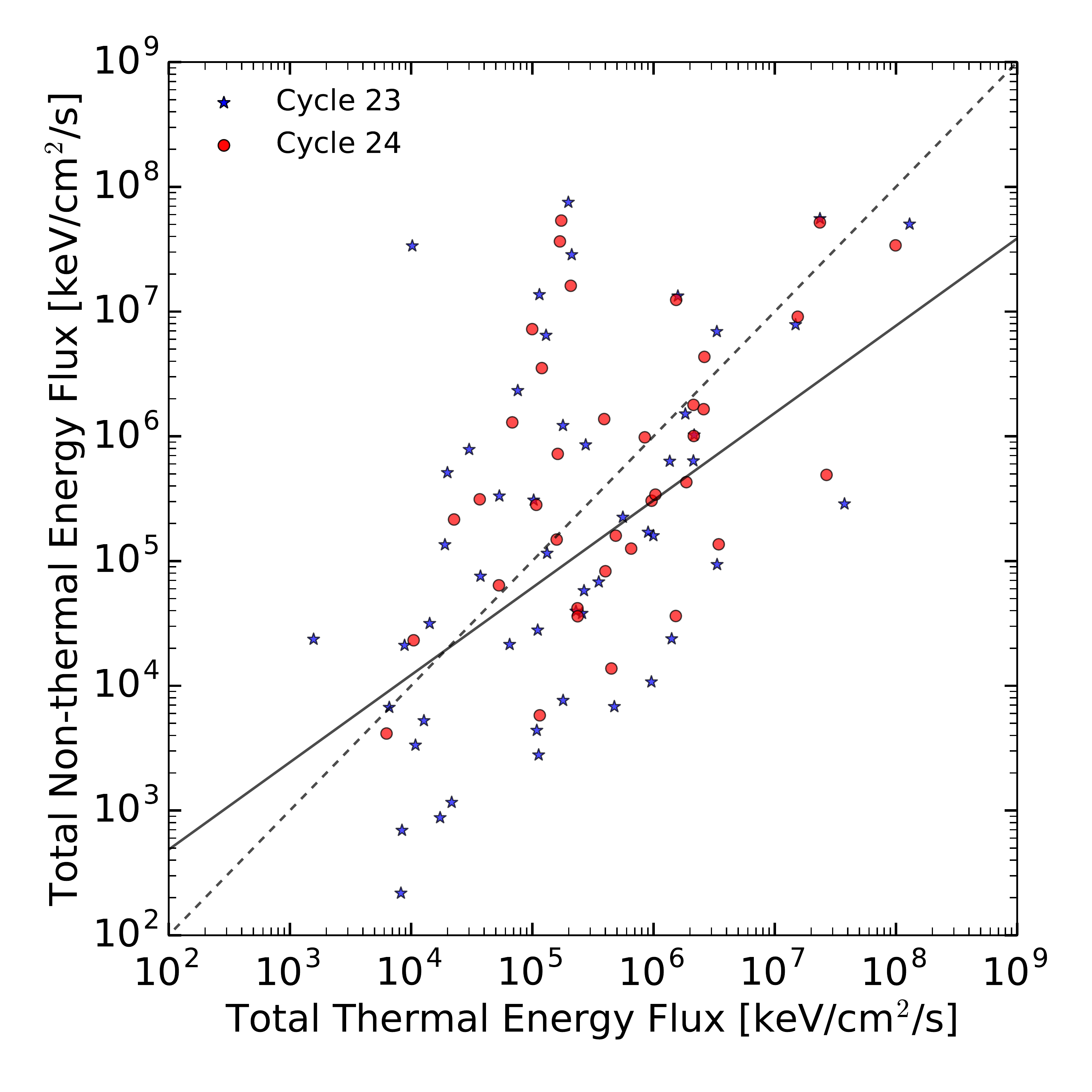}
\caption{Relation between the total thermal and non-thermal energy
  flux for the total of 90 flares (discarding all thermal events,
  cycle 23: blue stars, cycle 24: red circles). The solid line is a
  linear best fit to a power law with exponent 0.7 and the dashed
  line indicates equal thermal and non-thermal flux.}
\label{fig:thermal-nonthermal}
\end{figure}
A relatively direct way to estimate the importance of the non-thermal
emission is to calculate the total energy flux of the thermal
(bremsstrahlung) and non-thermal (broken power-law) components of the
fits to the observed X-ray spectra.

The total thermal energy flux depends only on the emission measure
$EM$ and temperature $T$ of the electrons as:
\begin{eqnarray}
F_{\rm tot,th} &=& 3\cdot10^4 \,                  
\sqrt{\frac{T}{10^6\rm{MK}}}\left(\frac{EM}{10^{45}\rm{cm^{-3}}}\right)
\,\rm{keV/cm^2/s}\,,
\end{eqnarray}
while the total energy flux of a broken power law model (in the same
units and with low and high energy indexes 1.5 and $\gamma$) is
\begin{eqnarray}
F_{\rm tot,nth} &=& 450 \frac{\gamma-1.5}{\gamma -2} F_{\rm nth}(E_{\rm break})
                    \left(\frac{E_{\rm break}}{15 {\rm keV}}\right)^2\,,
\end{eqnarray}
where $F_{\rm nth}(E_{\rm break})$ is the photon number flux
(\#/cm$^2$/s/keV) at the break energy.

Figure~\ref{fig:thermal-nonthermal} shows the resulting relation
between these energy fluxes for the two flare samples. A correlation
(linear correlation coefficient 0.53) can be detected and there is a
rough equipartition between energy fluxes. Most electron flare
acceleration models starting with a thermal plasma lead to a
quasi-thermal plus a power law component \citep[see,
e.g.][]{Petrosian-Liu-2004}, with the first producing the thermal and
the latter the non-thermal X-rays.  Since the bremsstrahlung yield is
primarily proportional to the average electron energy ($\propto kT$
and $\propto E_{break}$ for the two components, respectively), and
because $kT\sim E_{break}$, we expect a similar relation between the
two accelerated components as that between the two photon components.

\subsection{Time scales and thin target emission}
\begin{figure}[t]
\centering
\includegraphics[width=0.5\textwidth]{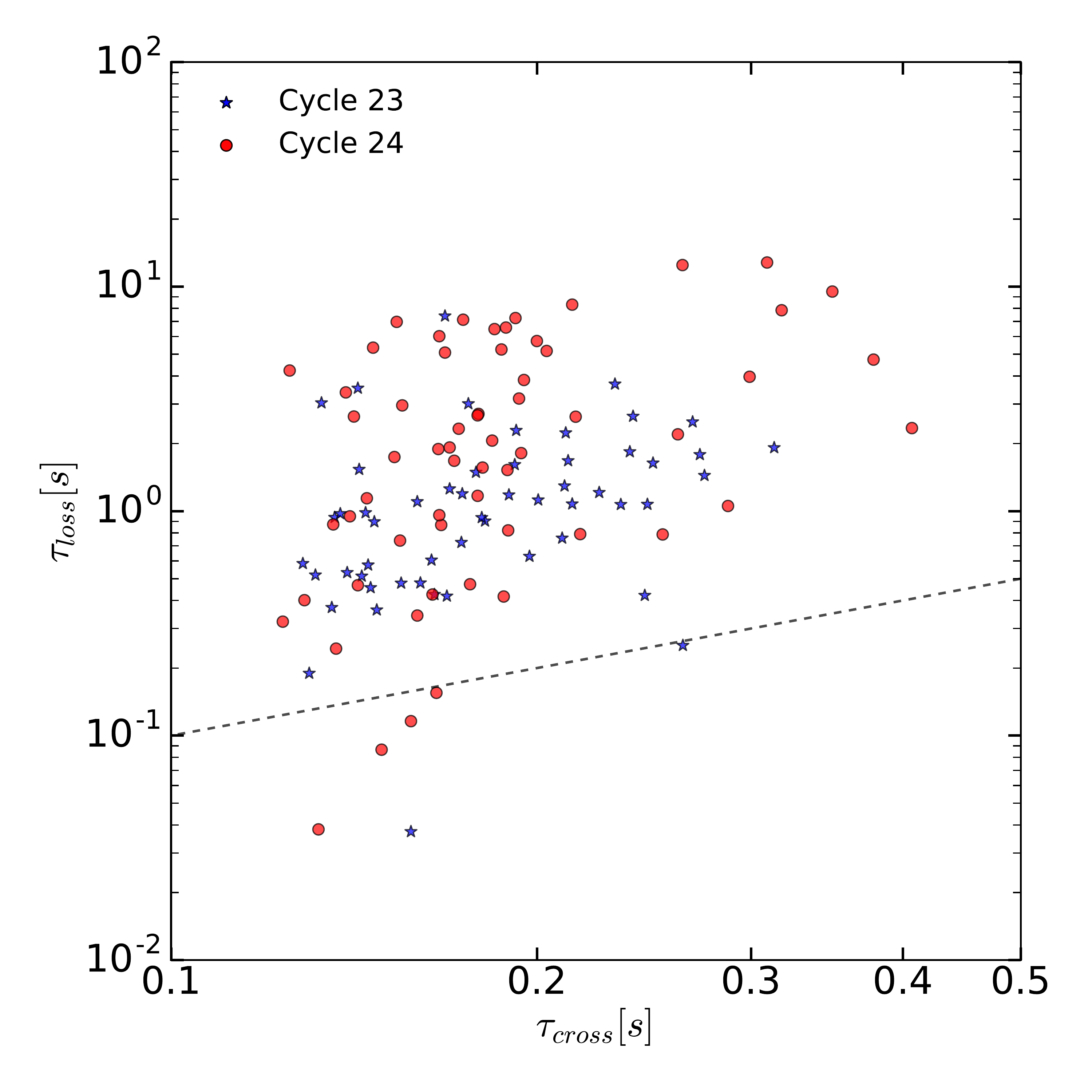}
\caption{Crossing time ($\tau_{cross}$) and energy loss time
  ($\tau_{loss}$) as calculated from the areas and densities found for
  each flare (cycle 23: blue stars, cycle 24: red circles) for 15~keV
  electrons. Most flares are above the dashed line of equal time
  scale.}
\label{fig:timescales}
\end{figure}
A thin-target model is the correct description for coronal source
emission if the time spent by the electrons in the source region is
shorter than the energy loss time (mainly due to elastic Coulomb
collisions at the non-relativistic energies under consideration here)
\begin{equation}
  \tau_L=\frac{E}{\dot{E}_L} = \frac{\sqrt{2}E^{3/2}}{4\pi r_0^2 n_e c
    \ln{\Lambda}} \, ,
\end{equation}
where the electron energy $E$ is measured in units of $mc^2$,
$r_0=2.8\cdot 10^{13}$~cm is the classical electron radius and
$\ln \Lambda\sim 20$ is the Coulomb logarithm. In absence of field
convergence or scattering the time spent in the source or the escape
time from it is equal to the time for crossing the source
$T_{\rm esc}\sim \tau_{\rm cross}=L/v$, for a source size $L$ and
electron velocity $v$.  We calculated the ambient electron density
$n_e$ for each flare from the emission measure of the thermal
component in the broken power-law fit ($n_e\sim\sqrt{EM/V}$) assuming
a filling factor of unity and a spherical source of $V\sim A^{2/3}$,
where $A\sim L^2$ is the projected area of the $50\%$ image contour.

Figure \ref{fig:timescales} compares the resulting two time scales for
our sample of flares at energy $E=15$ keV, close to the average
$E_{break}$. As evident, the thin-target assumption is justified down
to break energies for nearly all flares. Note that the distribution in
the figure would shift up (down) for higher (lower) electron energies
due to the energy dependences in $\tau_{\rm cross}$ and $\tau_L$. On
average, the two times become comparable at energies below 5 keV where
the thin-target assumption would breakdown.  In general, there can be
some trapping of the electrons so that the time they spend at the loop
top before escaping to the footpoints,
$\tau_{\rm esc}>\tau_{\rm cross}$.  This can come about if the field
lines converge toward lower heights or there is a scattering
agent. Coulomb collisions cannot be this agent because the Coulomb
scattering time is comparable to the loss time and therefore its
effect will be negligible. Turbulence can provide the
scattering. We can only have a low level of turbulence, so that
$\tau_{\rm esc}<\tau_L$. This appears, for example, to be the case for
two non-occulted flares studied by \citet{Chen-Petrosian-2013}. Using
an inversion technique they obtain all the above time scales plus the
acceleration and scattering time by turbulence and show that above
this condition is satisfied above $\sim 10$ keV and the collision loss
time is even longer than acceleration and energy diffusion time above
30 keV.

\subsection{The thin-target kappa model}
Assuming a thin-target situation we fit the spectra to that expected
from electrons with a kappa distribution. Since this distribution
contains a quasi-thermal component, we tested several fitting
methods. First we fit the spectra to a pure kappa function. We find
reasonable fits for most of the flares. But, in general, the $\chi^2$
(per free parameters) resulting from this model are higher than those
obtained by adding a separate thermal component. Also, as can be
expected, the resulting $\kappa$ values are systematically larger for
a pure kappa model.  Moreover, whenever there is a clear kink
separating the thermal and non-thermal energies, the kappa
distribution, with a weaker thermal part, especially for low $\kappa$,
is not able to properly fit all energy ranges, leading to large
residuals at high energies. These cases might indicate that a single
thin-target kappa model is not a viable model for many solar flares or
that there are multiple distinct plasma populations that are getting
combined in the integrated \emph{RHESSI} spectroscopy. This issue was
as already mentioned in KL2008 and followed up in the context of kappa
distributions in \citet{Oka-etal-2013}. In general, we found that for
some flares the broken power-law fit adjusts better to the spectra
than the kappa distribution.

\begin{figure}[t]
\centering
\includegraphics[width=0.5\textwidth]{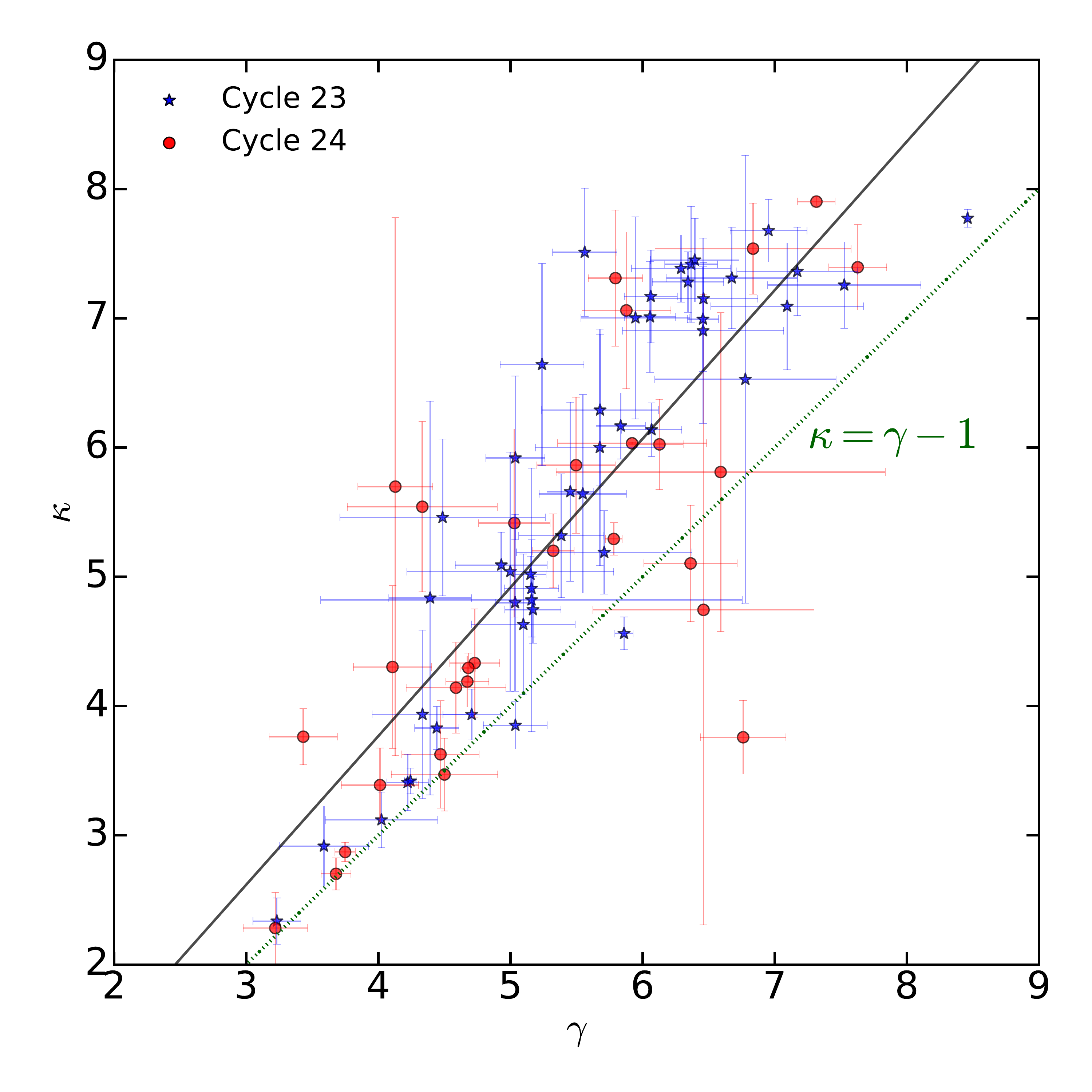}
\caption{Correlation between the broken power-law photon spectral
  index $\gamma$ and the electron $\kappa$ index values of the thermal
  plus thin-target kappa fit for the data of both solar cycles (cycle
  23: blue stars, cycle 24: red circles). The black dashed line shows
  the result of a linear fit to the complete dataset
  ($\kappa = 1.15\,\gamma - 0.83$).  The green dotted line indicates
  the theoretical thin target relation.}
\label{fig:corr_gamma_kappa}
\end{figure}
In Figure~\ref{fig:corr_gamma_kappa} we show a plot of $\kappa$ versus
the photon index $\gamma$.%
\footnote{Note that each $\kappa$ value has been estimated as the
  averaged result from all the individual fits from the single
  detectors (discarding detectors with different behaviors) and the
  error bars associated to each value have been estimated as the
  standard deviation amongst the different detector fits.}  As evident
from Equation (2) \citep[see also][]{Oka-etal-2013}, the relation
between the electron {\it flux} spectral index $\delta$ and the
electron kappa distribution index $\kappa$ is just $\delta=\kappa$. As
is well known for the thin-target case, we expect the relation
$\delta=\gamma-1$, resulting in $\kappa=\gamma-1$. This is shown by
the green dotted line in Figure~\ref{fig:corr_gamma_kappa}. The
linear-least-square fit to the whole dataset gives
$\kappa = 1.15\,\gamma - 0.83$ (black-dashed line) which is closer to
the expected thin-target relation, particularly for events with clear
non-thermal emission signal, i.e. small values of $\gamma$. The
discrepancies with the thin-target relation are larger for steep
($\gamma\gtrsim5$) spectra, which can be attributed to the larger
uncertainties when the thermal and non-thermal distribution are
difficult to separate. We emphasize that the deviation towards larger
than expected values of kappa, based on the thin-target model and
relation, mainly indicates the deviations between the two fitting
approaches, and cannot support independently a possible thick-target
regime ({\it cf.} the discussion of time scales in the previous
section).

\subsection{Temporal correlations and the Neupert effect}
A large fraction of flares in our sample shows a good correlation
between the derivative of {\it GOES} soft X-rays and the \emph{RHESSI}
hard X-ray emission. A good correlation generally implies an absence
of substantial lag. However, when there is a lag between the two light
curves it tends towards positive values, i.e. an earlier rise in {\it
  GOES} derivative.

The \citet{Neupert-1968} effect is a temporal correlation between the
soft X-ray flux and the the integral of the microwave flux. The
underlying physics of this effect is based on an energy argument
\citep[e.g.,][]{Li-etal-1993, Veronig-etal-2002, Veronig-etal-2005,
  Ning-Cao-2010} that the latter depends on the instantaneous flux of
non-thermal electrons that deposit their energy (by collisional
heating) to the dense chromosphere and drive evaporation of hot plasma
that emits soft X-rays. The time integration of the instantaneous
energy deposition rate equals the total energy deposited to the
thermal plasma, which is reflected in its enhanced temperature and
emission measure and thus the soft X-ray irradiance. Therefore, the
time derivative of the soft X-ray flux is expected to be correlated
with the instantaneous hard X-ray flux. Since the energy deposited by
accelerated electrons is related to the observed hard X-ray emission
via the escape time $T_{esc}$, we have as a restriction that we can
only observe the simple Neupert effect as long as $T_{esc}$ is
independent of time.

As pointed out by \citet{Liu-etal-2006}, neither the
non-thermal bremsstrahlung X-ray emission is a {\it linear} function of
the collisional energy deposition rate, nor the thermal X-ray emission
depends linearly on the total energy content of the thermal
plasma. Therefore, a perfect linear correlation for the Neupert effect
is not expected. In addition, the above energy argument is based on
the assumption that non-thermal electrons are the sole agent of energy
transport from the coronal loop-top to the footpoints. This is not
necessarily the case either, because other mechanisms, including
thermal conduction, can play some role and cause further deviations of
the temporal correlation \citep[e.g.\ ][found the ratio between
non-thermal to thermal energies to increases with flare
duration]{Saint-Hilaire-Benz-2005}.  In particular, the slightly
preferential positive lag of the hard X-ray flux from the \emph{GOES}
derivative suggests that in those flares substantial soft X-ray
emitting thermal plasma is present prior to the acceleration of
non-thermal electrons, which implies non-collisional heating
mechanisms such as thermal conduction \citep{Zarro-Lemen-1988,
  Battaglia-Fletcher-2009} or turbulence
\citep{Petrosian-Liu-2004}. On the other hand, the existence of
negative lags for a small fraction of the flares is consistent with
the finding of \citet{Liu-etal-2006}, who ascribed this to
hydrodynamic timescales for the deposited non-thermal energy to drive
chromospheric evaporation of thermal plasma. 

Despite the expectation that the partial occultation of soft X-ray
emitting plasma and footpoint hard X-rays can potentially cause
further deviations from the perfect Neupert effect compared with
on-disk flares, we do find strong correlations. This provides
additional support for the scenario that the primary particle
acceleration site is at or near the coronal X-ray source, rather than
at the chromospheric footprints \citep[e.g.,][]{Fletcher-Hudson-2008}.

\subsection{Source Morphologies}
As mentioned in Section~\ref{Sect:image}, there is no clear trend
towards positive or negative separations between low and high energy
source positions. The average separation is $d_{max}$=0.3~Mm, which is
not statistically significant. This result is in agreement with the
previous findings by KL2008. The trend is, however, inconsistent with
several individual case studies of coronal X-ray sources, which found,
in general, that the higher energy emission is located at greater
heights
\citep[e.g.,][]{Sui-Holman-2003,Sui-etal-2004,Liu-etal-2004,Liu-etal-2008,Liu-etal-2009}. We
found in certain flares of our sample that there are multiple sources
above the loop-top (see e.g Figure~\ref{fig:clean_im}). More detailed
studies of such events can be found in \citet{Liu-etal-2013},
\citet{Krucker-Battaglia-2014}, \citet{Oka-etal-2015}, and
\citet{Effenberger-etal-2016}. A few flares showed also high-energy
emission coming from regions lower in the corona than the location of
the low-energy centroid. A possible explanation for this feature is a
situation in which the flare loop is very occulted, so that only the
above-the-looptop sources are visible. In this situation, the high
energy emission is closer to the reconnection region and thus to the
limb \citep{Liu-etal-2013, Oka-etal-2015}.

\clearpage
\section{Summary}
We have analyzed X-ray light curves, images and spectra of 116
partially occulted flares during solar cycle 23 and 24. The additional
availability of \emph{SDO} and \emph{STEREO} observations during cycle
24 allowed for supplementary information to characterize the limb
flares. EUV observations from \emph{STEREO} allowed us to estimate the
actual \goes classification and the high-cadence AIA images helped to
confirm the actual occultation of the flare with greater confidence
and provided valuable context for the interpretation of the individual
flare evolution. From \emph{STEREO} we further obtained the position
of the active region and determined the height of the coronal
source. Our results can be summarized as follows:

\begin{enumerate}
\item We found no significant difference in the statistics of the
  derived flare properties between the two cycles.

\item We use a thermal plus a power-law (with index $\gamma$) model to
  describe the \rhessi X-ray spectra. Most flares are dominated by the
  thermal component and about 20\% show no discernible non-thermal
  part. From the spectral fitting parameters (emission measure,
  temperature, break energy and power-law spectral indexes) we
  compared the emitted energy fluxes (during 20 second around the impulsive
  phase) in thermal and non-thermal photons. We found some correlation
  and comparable energies. Using the density of the thermal component
  (derived from the emission measure and size of the source) we found
  that the energy loss time is much longer than the free source
  crossing time indicating that we are dealing with a thin-target
  model.

\item We also fitted the photon spectra by a thin-target
  bremsstrahlung emission model from electrons with a {\it kappa}
  distribution, which consists of a Maxwellian plus a non-thermal
  component (with high energy index $\kappa$).  Although reasonable
  fits are obtained in this procedure, we found that the electrons in
  the Maxwellian part often cannot describe the prominent thermal
  component of the photon spectra adequately, and that better fits can
  be obtained by the addition of a more prominent thermal source. We
  found that the spectra of occulted flares tend to be softer than
  general disk flares with the relation between the photon and
  electron indexes, $\kappa=1.18\gamma -0.84$ to be in rough agreement
  with that expected in a thin-target model at lower values for the
  spectral index. It deviates significantly from this relation for
  high values of the indexes where the spectra are dominated by the
  thermal component and errors are large.

\item We found no trend for large spatial separations between low and
  high energy hard X-ray components in the spectral images of our
  sample. There are, however, notable exceptions with larger
  separations and a richer coronal source structure. The estimations
  of source height and \emph{GOES} classification with \emph{STEREO}
  observations reveal a large variety of coronal source positions with
  heights up to 52~Mm and differences in \emph{GOES} class showing for
  example a strong C class flare to be actually a X class flare.

\item We found a significant correlation between the time derivative
  of the soft X-ray and the observed hard X-rays light curves for a
  large fraction of our sample, with a mean lag time of near zero,
  consistent with earlier studies for on-disk flares
  \citep{Veronig-etal-2002}. This confirms the presence of the simple
  Neupert effect for purely coronal sources and supports the scenario
  that the main source of non-thermal particles is produced near the
  looptop. The lags found in some flares indicate that additional
  processes like thermal conduction can play important roles.

\end{enumerate}

As mentioned at the outset, partially occulted flares give us direct
information on the physical conditions at the acceleration site near
the loop-top sources and the spectrum of the accelerated
electrons. These can be used to put meaningful constrains on the
characteristics of the acceleration mechanisms.
 
The STIX X-ray instrument on the upcoming \emph{Solar Orbiter} mission
will provide further capabilities to investigate coronal sources of
flares from multiple perspectives, allowing us to observe and model the
entire flare with greater detail. Hard X-ray focusing optics like
\emph{FOXSI} \citep{Krucker-etal-2013,Krucker-etal-2014} can provide
simultaneous imaging of the chromospheric footpoints and coronal
sources from a single view point due to much improved dynamic range
compared to \emph{RHESSI} and will thus enable additional insights into
electron acceleration and transport processes in the corona, even for
on-disk flares.

\acknowledgements Work performed by F.E., F.R.dC.\ and V.P.\ is
supported by NASA grants NNX13AF79G and NNX14AG03G. MO was supported
by NASA grants NNX08AO83G at UC Berkeley and NNH16AC60I at Los Alamos
National Laboratory (LANL). W.L.\ was supported by NASA HGI grant
NNX16AF78G. L.G.\ was supported by an NSF Faculty Development Grant
(AGS-1429512) to the University of Minnesota. Work performed by
F.R.dC.\ and W.L.\ is part of the team effort under the title of
``Diagnosing heating mechanisms in solar flares through spectroscopic
observations of flare ribbons'' at the International Space Science
Institute (ISSI). M.O. and F.E. are part of the ISSI team ``Particle
Acceleration in Solar Flares and Terrestrial Substorms''.

\newpage
\begin{appendix}\label{Sect:appendix}
  Here we reproduce the analysis of the flares from cycle 23 in the
  KL2008 sample using our analysis procedures. Note that flare 43
  (October 23, 2003) showed footpoint-like emission in our images and
  was thus discarded for this study.

\begin{deluxetable*}{rllcrr|ccc|ccc|c|cr}
\tabletypesize{\scriptsize}
\tablecolumns{15} 
\tablewidth{0pt}
\tablenum{2}
\tablecaption{\label{KL_list}Analysis results for partially occulted flares from KL2008}
\tablehead{
  \colhead{\#} & 
  \colhead{Date} & 
  \colhead{Time} &
  \colhead{{\it GOES}} &
  \colhead{Sol-X} &
  \colhead{Sol-Y} &
  \colhead{$T_{th}$} &
  \colhead{$E_{break}$} &
  \colhead{$\gamma$} &
  \colhead{$T_{th}^{\kappa}$} &
  \colhead{$T^{\kappa}$} &
  \colhead{$\kappa$} &
  \colhead{$d_{max}$\tablenotemark{(a)}} &
  \colhead{Lin.}  &
  \colhead{Lag \tablenotemark{(b)}}\vspace{0.1cm}\\
  \colhead{} &
  \colhead{} &
  \colhead{(UT)} &
  \colhead{Class} &
  \colhead{(arcs.)} &
  \colhead{(arcs.)} &
  \colhead{(MK)} &
  \colhead{(keV)} &      
  \colhead{} &
  \colhead{(MK)} &
  \colhead{(MK)} &
  \colhead{} &
  \colhead{(Mm)} &
  \colhead{} &
  \colhead{(s)}\vspace{0.1cm}\\
  \colhead{1} &
  \colhead{2} &
  \colhead{3} &
  \colhead{4} &
  \colhead{5} &
  \colhead{6} &
  \colhead{7} &
  \colhead{8} &
  \colhead{9} &
  \colhead{10} &      
  \colhead{11} &
  \colhead{12} &
  \colhead{13} &
  \colhead{14} &
  \colhead{15}}
\startdata
1  &  2002 Mar 07 &  17:50:44 &  C2.5 &    -961.5 &   -176.4 &      21.4 &        11.6 &     5.86 &               32.0 &         10.1 &     4.56 &           -0.4 &        0.77 &  12 \\
2  &  2002 Mar 28 &  17:56:07 &  C7.6 &     965.7 &    -64.2 &      21.6 &        17.6 &     5.55 &               15.6 &         12.9 &     5.64 &           -0.6 &        0.66 &  -4 \\
3  &  2002 Apr 04 &  10:43:52 &  M1.4 &    -896.8 &   -347.5 &      25.4 &        18.8 &     5.16 &               16.9 &         11.0 &     4.91 &            0.5 &        0.86 &   0 \\
4  &  2002 Apr 04 &  15:29:14 &  M6.1 &    -909.3 &   -333.8 &      27.0 &        19.6 &     5.17 &               26.5 &          9.8 &     4.75 &            0.7 &        0.87 &   4 \\
5  &  2002 Apr 18 &  06:52:04 &  C9.4 &     913.7 &    314.1 &      19.9 &        17.0 &     5.38 &               18.3 &          9.9 &     5.32 &            3.7 &         - & - \\
6  &  2002 Apr 22 &  12:05:56 &  C2.8 &     900.2 &   -319.3 &      20.9 &        15.8 &     4.02 &               20.6 &          7.6 &     3.12 &            3.5 &         - & - \\
7  &  2002 Apr 29 &  13:00:55 &  C2.2 &    -904.5 &   -323.6 &     10.1 &        13.3 &     8.06 &                - &              - &      - &            0.0 &         - & - \\
8  &  2002 Apr 30 &  00:32:48 &  C7.8 &    -882.8 &   -373.7 &      20.9 &        18.6 &     5.68 &               18.6 &          7.1 &     6.00 &            0.5 &         - & - \\
9  &  2002 Apr 30 &  08:20:44 &  M1.3 &    -891.5 &   -357.5 &      23.0 &        17.8 &     5.15 &               20.2 &         10.3 &     5.02 &            1.4 &        0.87 &   0 \\
10 &  2002 May 17 &  02:01:28 &  C5.1 &    -929.5 &    227.5 &      26.0 &        17.5 &     4.49 &               19.3 &         11.3 &     5.46 &            0.0 &         - & - \\
11 &  2002 May 17 &  07:32:40 &  M1.5 &    -931.2 &    230.2 &      26.6 &        19.9 &     6.06 &               17.3 &         17.5 &     7.17 &            1.1 &        0.47 &  -8 \\
12 &  2002 Jul 05 &  08:03:06 &  C7.8 &     917.3 &   -291.2 &      20.8 &        19.1 &     7.53 &               19.1 &         11.0 &     7.26 &            0.5 &         - & - \\
13 &  2002 Jul 06 &  03:32:15 &  M1.8 &     911.6 &   -281.4 &      24.9 &        18.9 &     5.56 &               21.7 &         31.8 &     7.51 &            2.1 &         - & - \\
14 &  2002 Jul 08 &  09:15:39 &  M1.6 &    -891.8 &    330.5 &      22.8 &        18.1 &     5.45 &               17.6 &         12.6 &     5.66 &            1.6 &        0.52 &   4 \\
15 &  2002 Jul 09 &  04:03:28 &  C8.6 &     896.3 &    322.4 &      20.9 &         - &      - &                - &          - &      - &            - &        0.60 &  24 \\
16 &  2002 Aug 04 &  09:38:51 &  M6.6 &     945.1 &   -338.0 &      21.0 &        18.8 &     8.46 &               23.9 &         12.0 &     7.77 &           -0.9 &         - & - \\
17 &  2002 Aug 04 &  14:14:48 &  C6.9 &     909.5 &   -310.7 &      23.2 &        18.3 &     6.95 &               15.2 &         14.5 &     7.68 &            1.3 &        0.66 &   0 \\
18 &  2002 Aug 28 &  18:54:44 &  M4.6 &    -937.9 &    158.3 &      32.5 &        19.0 &     4.44 &               32.9 &          9.8 &     3.83 &            0.1 &        0.60 &  12 \\
19 &  2002 Aug 28 &  21:43:23 &  M1.1 &     845.4 &   -445.7 &      26.7 &        19.4 &     5.04 &               26.1 &          8.8 &     4.80 &            0.7 &        0.49 &   0 \\
20 &  2002 Aug 29 &  02:50:28 &  M1.6 &    -946.9 &    141.5 &      26.2 &        20.2 &     5.83 &               20.6 &          9.9 &     6.17 &            1.0 &        0.74 &   8 \\
21 &  2002 Aug 29 &  05:43:56 &  C9.2 &     840.3 &    449.9 &      33.7 &        21.9 &     5.24 &               28.5 &         23.1 &     6.64 &            0.0 &        0.88 &   4 \\
22 &  2002 Sep 06 &  16:27:00 &  C9.2 &    -954.6 &   -101.7 &      23.5 &        16.7 &     4.22 &               22.4 &          6.4 &     3.41 &           -1.7 &        0.88 &   0 \\
23 &  2002 Oct 16 &  15:57:20 &  C6.5 &    -830.4 &    499.6 &      23.8 &        17.9 &     6.78 &               17.9 &         21.7 &     6.53 &           -1.5 &         - & - \\
24 &  2002 Nov 15 &  01:08:36 &  M2.4 &    -931.9 &   -290.0 &      24.5 &        18.4 &     5.95 &               14.9 &         14.7 &     7.00 &            2.1 &        0.93 &   0 \\
25 &  2002 Nov 23 &  01:21:40 &  C2.1 &    -946.6 &    234.5 &      22.0 &        16.6 &     5.00 &               21.0 &         13.2 &     5.04 &            0.9 &         - & - \\
26 &  2003 Jan 21 &  01:25:51 &  C2.0 &    -935.8 &   -299.2 &      31.2 &         - &      - &                - &          - &      - &            - &         - & - \\
27 &  2003 Feb 01 &  08:57:28 &  M1.2 &    -966.4 &   -251.1 &      25.3 &        18.1 &     6.37 &               16.2 &         16.4 &     7.42 &           7.7 &        0.86 &   4 \\
28 &  2003 Feb 01 &  19:41:33 &  C9.9 &    -965.3 &   -237.4 &      33.9 &        16.6 &     7.39 &                - &          - &      - &            0.5 &         - & - \\
29 &  2003 Feb 14 &  09:16:15 &  M1.2 &     955.0 &    207.1 &      24.5 &        19.1 &     5.68 &               21.6 &         10.5 &     6.29 &            0.0 &        0.76 &   4 \\
30 &  2003 Mar 27 &  14:52:04 &  C2.3 &     925.5 &    299.5 &      27.4 &        11.2 &     4.39 &               23.2 &         17.2 &     4.84 &           -0.7 &        0.64 &  16 \\
31 &  2003 Apr 24 &  04:53:44 &  C7.1 &     925.3 &    279.9 &      20.9 &        17.2 &     6.06 &               14.3 &         16.0 &     7.01 &           -0.7 &        0.83 &   0 \\
32 &  2003 Apr 24 &  06:35:08 &  C1.0 &     907.6 &    267.9 &      29.6 &        17.2 &     3.59 &               19.9 &         12.5 &     2.92 &           -0.7 &        0.88 &   0 \\
33 &  2003 May 07 &  20:47:48 &  C5.9 &    -916.8 &    263.8 &      22.8 &        19.0 &     7.09 &               18.9 &         12.9 &     7.09 &            2.5 &        0.94 &   0 \\
34 &  2003 Jun 01 &  12:48:32 &  M1.0 &    -936.6 &    172.3 &      28.7 &        20.4 &     4.93 &               19.3 &          9.4 &     5.09 &           -1.0 &        0.87 &  -8 \\
35 &  2003 Jun 02 &  08:15:12 &  M3.9 &     940.1 &   -137.5 &      22.6 &        17.2 &     5.04 &               18.1 &         12.7 &     5.92 &            0.0 &         - & - \\
36 &  2003 Jun 06 &  16:17:10 &  C2.5 &    -920.2 &   -243.6 &      17.5 &        13.4 &     8.35 &                - &          - &      - &           -0.3 &        0.76 &  12 \\
37 &  2003 Aug 21 &  15:19:02 &  C4.9 &     940.4 &   -178.6 &      22.9 &        16.9 &     4.33 &               21.6 &          9.0 &     3.94 &            0.1 &         - & - \\
38 &  2003 Sep 15 &  16:38:51 &  C1.5 &     953.6 &   -122.4 &      22.4 &        13.1 &     6.54 &                - &          - &      - &           -0.9 &        0.70 &   0 \\
39 &  2003 Sep 29 &  16:09:07 &  C5.1 &    -953.8 &   -136.6 &      20.9 &        17.2 &     6.68 &               18.9 &         14.5 &     7.31 &           -1.7 &        0.66 &   4 \\
40 &  2003 Oct 21 &  23:07:04 &  M2.4 &    -945.8 &   -277.8 &      26.0 &        20.0 &     6.46 &               12.1 &         11.0 &     7.15 &            2.5 &         - & - \\
41 &  2003 Oct 22 &  18:43:08 &  C6.0 &    -937.2 &   -285.1 &      24.5 &        17.2 &     7.17 &               26.1 &         11.4 &     7.36 &            0.3 &         - & - \\
42 &  2003 Oct 22 &  21:56:24 &  M2.1 &    -943.0 &   -284.7 &      23.6 &        17.4 &     6.40 &               15.3 &         17.3 &     7.45 &           -0.5 &        0.83 &   0 \\
43 &  2003 Oct 23 &  01:06:24 &  C4.0 &    -938.8 &   -291.9 &       - &         - &      - &                - &          - &      - &            - &         - & - \\
44 &  2003 Nov 04 &  14:47:28 &  C7.5 &     967.9 &    161.7 &      24.5 &        17.6 &     6.29 &               11.3 &         18.5 &     7.39 &           -1.7 &         - & - \\
45 &  2003 Nov 04 &  15:30:08 &  C5.6 &     956.5 &    187.4 &      18.9 &        17.1 &     4.71 &               15.7 &          8.3 &     3.93 &            0.2 &        0.47 &  24 \\
46 &  2003 Nov 05 &  01:58:53 &  C7.2 &     965.8 &    170.0 &      28.7 &        16.3 &     5.10 &               28.2 &          9.6 &     4.63 &           -1.5 &         - & - \\
47 &  2003 Nov 18 &  09:43:58 &  M4.5 &   -1002.4 &   -232.2 &      16.5 &        15.2 &     6.46 &               15.5 &         11.8 &     6.99 &          19.7 &         - & - \\
48 &  2003 Nov 18 &  22:17:41 &  C6.1 &    -939.6 &   -271.2 &      29.4 &        14.9 &     3.23 &               20.4 &         15.9 &     2.34 &           -0.3 &        0.71 &   0 \\
49 &  2003 Nov 19 &  10:07:18 &  C2.7 &    -931.5 &   -272.9 &      13.2 &        11.1 &     5.04 &               11.5 &          0.3 &     3.85 &           -0.6 &        0.78 &   8 \\
50 &  2004 Mar 05 &  08:54:21 &  C6.6 &    -948.0 &   -282.7 &      29.1 &        20.5 &     5.16 &               20.7 &         11.5 &     4.82 &           -1.4 &        0.35 &   8 \\
51 &  2004 Mar 24 &  23:25:22 &  M1.5 &    -941.1 &    245.6 &      24.4 &        19.3 &     6.34 &               15.6 &         12.7 &     7.28 &            0.0 &        0.62 &  20 \\
52 &  2004 Jul 15 &  22:27:05 &  C7.9 &    -944.0 &    162.6 &      29.8 &        19.5 &     6.46 &               30.7 &         22.9 &     6.90 &           -1.3 &        0.91 &   0 \\
53 &  2004 Jul 17 &  03:46:25 &  C4.2 &    -930.5 &    146.2 &      22.0 &        17.3 &     4.24 &               15.6 &         10.5 &     3.42 &           -0.9 &        0.81 &  16 \\
54 &  2004 Aug 18 &  08:41:05 &  C6.1 &     936.2 &   -212.2 &      27.2 &        18.8 &     5.71 &               11.7 &         12.5 &     5.19 &           -0.6 &        0.61 &  -4 \\
55 &  2004 Aug 19 &  06:54:04 &  M3.0 &     943.0 &   -213.2 &      28.4 &        22.6 &     6.07 &               17.6 &         10.2 &     6.14 &            0.7 &        0.90 &   0 \\
\enddata
\tablenotetext{(a)}{\, A positive $d_{max}$ implies a high-energy source at
greater radial distance.}
\tablenotetext{(b)}{\, Positive lags indicate a delay in the \emph{RHESSI}
  light curve with respect to the \emph{GOES} soft X-ray derivative.}
\end{deluxetable*}

\end{appendix}

\newpage
\bibliography{references}
\end{document}